\documentclass[]{aa}  

\usepackage{graphicx}
\usepackage{txfonts}
\usepackage{hyperref}
\usepackage{upgreek}
\usepackage{comment}
\usepackage{mathtools}
\usepackage{ulem}

\newcommand{\sectionname}{Sect.}
\newcommand{\dd}{\textrm{d}}

\newcommand{\deriv} [2] {\frac {\textrm{d} #1 } {\textrm{d} #2} }

\newcommand{\algn} [1] {
\begin{align} #1
\end{align}}

\let\originaleqref\eqref
\renewcommand{\eqref}{Eq.~\originaleqref}
\newcommand{\eq}[1] {Eq.\,(\ref{#1})}
\newcommand{\eqss}[2]{Eqs.~(\ref{#1})-(\ref{#2})}
\makeatletter
\newcommand{\eqs}[1]{%
    Eqs.~(\ref{#1})\checknextarg}
\newcommand{\checknextarg}{\@ifnextchar\bgroup{\gobblenextarg}{}}
\newcommand{\gobblenextarg}[1]{\@ifnextchar\bgroup{, (\ref{#1})\gobblenextarg}{ and (\ref{#1})}}
\makeatother

\begin{document} 


   \title{Amplitude of solar gravity modes generated by penetrative plumes}

   \subtitle{}
  \titlerunning{Plume-induced asymptotic g-modes} 
  
   \author{C. Pin\c con\inst{1}, T. Appourchaux\inst{2} \and G. Buldgen\inst{3}
          }
          
  \authorrunning{C. Pin\c con et al.}
   \institute{STAR Institute, Université de Liège, 19C Allée du 6 Août, B-4000 Liège, Belgium
         \and
           Université Paris-Sud, Institut d’Astrophysique Spatiale, UMR 8617, CNRS, Bâtiment 121, 91405 Orsay Cedex, France
          \and
          Observatoire de Genève, Université de Genève, 51 Ch. Des Maillettes, CH-1290 Sauverny, Suisse
             }

   \date{\today}

  \abstract
  {The observation of gravity modes is expected to give us unprecedented insights into the inner dynamics of the Sun. Nevertheless, there is currently no consensus on their detection. Within this framework, predicting their amplitudes is essential to guide future observational strategies and seismic studies.}
  {While previous estimates considered convective turbulent eddies as the driving mechanism, we aim to predict the amplitude of low-frequency asymptotic gravity modes generated by penetrative convection at the top of the radiative zone.}
   {A generation model previously developed for progressive gravity waves is adapted to the case of resonant gravity modes. The stellar oscillation equations are analyzed considering the plume ram pressure at the top of the radiative zone as the forcing term. The plume velocity field is modeled in an analytical form.}
   { We obtain an analytical expression for the mode energy. It is found to depend critically on the time evolution of the plumes inside the generation region. Using a solar model, we then compute the apparent surface radial velocity of low-degree gravity modes such as it would be measured by the GOLF instrument, in the frequency range $10~\mu$Hz$\le \nu \le 100~\mu$Hz. In case of a Gaussian plume time evolution, gravity modes turn out to be undetectable because of too small surface amplitudes. This holds true despite a wide range of values considered for the parameters of the model. In the other limiting case of an exponential time evolution, plumes are expected to drive gravity modes in a much more efficient way because of a much higher temporal coupling between the plumes and the modes than in the Gaussian case. Using reasonable values for the plume parameters based on semi-analytical models, the apparent surface velocities in this case turn out to be one order of magnitude smaller than the 22-years GOLF detection threshold and than the previous estimates considering turbulent pressure as the driving mechanism, with a maximum value of $0.05$~cm~s${}^{-1}$ for $\ell =1$ and $\nu\approx 100~\mu$Hz. When accounting for uncertainties on the plume parameters, the apparent surface velocities in the most favorable plausible case become comparable to those predicted with turbulent pressure, and the GOLF observation time required for a detection at \smash{$ \nu \approx100~\mu$Hz} and $\ell=1$ is reduced to about $50~$yrs.}
   {Penetrative convection can drive gravity modes in the most favorable plausible case as efficiently as turbulent pressure, with amplitudes slightly below the current detection threshold. When detected in the future, the measurement of their amplitudes is expected to provide infrmation on the plume dynamics at the base of the convective zone. In order to make a proper interpretation, this potential nevertheless requires further theoretical improvements in our description of penetrative plumes.} 


   \maketitle


\section{Introduction}

In the last decades, the study of global acoustic modes revealed precious information on the internal properties of the Sun \citep[e.g., see][for interesting reviews]{Thompson2003,Basu2008,Kosovichev2011,Buldgen2019,JCD2020}. Among noteworthy results, the sound speed and rotation rate could be probed with a high precision down to $0.08R_\odot$ and $0.2R_\odot$, respectively. Nevertheless, the study of acoustic modes hardly gives access to the properties of deeper layers, where most of the solar luminosity is produced. Besides, gravity modes, which propagate in the central radiative zone, have the potential to probe the micro and macrophysics inside the solar core and test further stellar models \citep[e.g.,][]{Appourchaux2010}. For instance, the detection of gravity modes, combined with neutrinos flux measurements, can be expected to improve our representation of nuclear reaction rates and electron screening \citep[e.g.,][]{Mussack2011,Lopes2014,Bonventre2018}. The measurement of their rotational splittings can also give the possibility to unveil the deep rotation profile and put stringent constraints on the angular momentum history of the Sun \citep[e.g.,][]{Eggenberger2019}. From a wider perspective, all the information brought by solar gravity modes can permit, on the one hand, to calibrate stellar models and help improve our understanding of the whole stellar evolution. On the other hand, potential discrepancies between theory and observations in the earlier or later evolutionary phases can also give evidence for a change of regimes in the internal processes at work along the star lifetime \citep[e.g.,][]{Gehan2018,Eggenberger2019b}.

The quest for the solar gravity modes began more than forty years ago. Still, gravity modes are evanescent in the convective envelope and have very small amplitudes at the solar surface; they are thus very difficult to observe. Their detection have been claimed by several studies \citep[e.g.,][]{Brookes1976,Severnyi1976,Delache1983,Thomson1995,TC2004,Garcia2007}, but none of these results has been independently confirmed \citep[e.g.,][]{Appourchaux2013}. The most recent claim of their detection is imputed to \cite{Fossat2017} and \cite{Fossat2018}. Using data of the GOLF instrument \citep[e.g.,][]{Gabriel1995}, the authors indirectly found evidence for the signature of low-frequency gravity modes in the time variations of the large frequency separation of high-frequency acoustic modes. Their analysis led to the identification of hundreds of gravity modes with angular degrees between $\ell=1$ and $\ell=4$, from which they could infer the asymptotic period-spacing and the mean core rotation rate. Although this work was subsequently reproduced by other studies, the robustness of the analysis was seriously put into question \citep{Schunker2018,Appourchaux2019}. Indeed, the result was shown to be very sensitive to the parameters of the time series considered in the analysis, as the start time, and somehow appears to be an artifact of the methodology. Theoretical studies on the coupling between acoustic and gravity modes reinforced the fragility of this result \citep{Scherrer2019,Boning2019}. At the time of writing, no robust detection of the solar gravity modes has therefore been unequivocally confirmed yet.

Within this framework, theoretical estimates of the amplitude of solar gravity modes are needful to help design future observational missions and guide future seismic studies. 
The amplitude of gravity modes, as acoustic modes, results from a balance between the driving by convective motions and damping processes \citep[e.g.,][]{Belkacem2011b}. Previous theoretical estimates mainly considered the Reynolds stress of turbulent convective eddies as the source mechanism, or stochastic excitation \citep[e.g.,][]{Gough1985,Kumar1996,Belkacem2009}. Guided by global 3D numerical simulations of the solar convective zone and using reasonable values for their model parameters, \cite{Belkacem2009} determined that the amplitudes of asymptotic gravity modes, that is, with high radial orders and oscillation frequencies between $10~\mu$Hz $\lesssim \nu  \lesssim 100~\mu$Hz, are likely to lie slightly below the current GOLF detection threshold. Because of the non-detection of gravity modes, these predictions set an upper constraint on the Sun convective velocity in the excitation region (i.e., around $0.8~R_\odot$). However, all these previous estimates did not account for the contribution from the penetration of convective plumes at the base of the convective region (or penetrative convection) to the mode driving.

Convective plumes are strong downdrafts originating from diving cool granules at the solar surface. They develop by turbulent entrainment of matter as coherent structures when crossing the convective region \citep[e.g.,][]{Turner1986,Rieutord1995}. As the plumes reach the bottom of the convective bulk, they can penetrate into the underlying stably-stratified radiative layers; there, the plumes are braked by buoyancy and can transfer a part of their kinetic energy into gravity waves. While this excitation mechanism is ubiquitous in numerical simulations of extended convective envelopes overlying radiative zones \citep[e.g.,][]{Andersen1996,Dintrans2005,Kiraga2005,Rogers2006b,Rogers2013,Alvan2014,Edelmann2019}, the covered values of the dimensionless control parameters are far from stellar regimes. Quantitative estimates by means of semi-analytical excitation models are thus required and complementary \citep[e.g.,][]{Rempel2004}. Motivated by the issue of the redistribution of angular momentum in stellar interiors, \cite{Pincon2016} modeled this driving process to predict the amplitude of very-low-frequency progressive gravity waves propagating in the radiative zone of the Sun. They demonstrated that this process generates low-frequency gravity waves more efficiently than turbulent pressure and can have an important impact on the angular momentum evolution of low-mass stars \citep{Pincon2017}. However, no application regarding the amplitude of the solar gravity modes with much higher frequencies has been undertaken so far.

In this work, we aim to estimate the amplitude of the solar gravity modes generated by penetrative convection following the model of \cite{Pincon2016}.
We limit the study to asymptotic gravity modes in the frequency range between $10~\mu$Hz and $100~\mu$Hz (i.e., with a number of radial nodes much higher than unity). The damping of such modes is dominated by the radiative losses and is analytically tractable in the considered frequency range under the quasi-adiabatic limit \citep[e.g.][]{Dziembowski1977b,Belkacem2009}. In contrast, at higher frequencies, the computation of the mode damping requires to account for the interaction between oscillations and convection, which is much more complex and out scope of this paper \citep[e.g.,][]{Belkacem2011b}. The paper is organized as follows. In \sectionname{}~\ref{generation model}, an analytical expression for the mode energy is derived based on the model of \cite{Pincon2016}. In \sectionname{}~\ref{apparent}, the apparent surface radial velocity of gravity modes is estimated from this expression for a solar model and is compared to the GOLF data detection threshold. The results are discussed in \sectionname{}~\ref{discussion}. Conclusions are formulated in \sectionname{}~\ref{conclusion}.

\section{Excitation model by penetrative convection}
\label{generation model}

In this section, we derive an analytical expression for the energy of gravity modes excited by penetrative convection. The asymptotic (i.e., short-wavelength) approximation is used and the quasi-adiabatic limit is considered (i.e., non-adiabatic effects are globally considered as small perturbations). Both approximations are justified for the Sun in the considered frequency range (i.e., between $10~\mu$Hz and $100~\mu$Hz). The detailed derivation steps and technical issues are described in \appendixname{}~\ref{derivation}.

\subsection{Oscillation equations forced by penetrative plumes}
\label{forced equation}

Following \cite{Pincon2016}, the velocity field in the stellar frame is decomposed into a component associated with the convective plumes and a perturbation associated with the gravity modes. The source term in the linearized momentum equation is assumed to be the ram pressure exerted by the ensemble of convective plumes at the top of the radiative region. The feedback from the oscillations on the plume structure and dynamics is neglected. In other words, we assume that the wave energy is much smaller than the plume kinetic energy at the base of the convection zone, which will be checked a posteriori. The effect of the Coriolis force on both the oscillations and the plumes is not considered. This is justified for the gravity modes as the solar rotation period is much lower than the modal periods. For the plumes, the effect of the buoyancy work is predominant so that the plume Rossby number is expected to be very low in the case of slow rotators as the Sun.

Within this framework, the forced linear non-adiabatic oscillation equation reads (see \appendixname{}~\ref{forced momentum} for details)
\algn{
\partial _t^2 \vec{\xi} +\vec{\mathcal{L}}^{\rm ad}\left( \vec{\xi}\right) +\vec{\mathcal{L}}^{\rm nad}\left(H \frac{\delta S}{c_p} \right)=-\frac{1}{\rho}\vec{\nabla} \cdot (\rho \vec{\mathcal{V}}_{\rm p}\otimes \vec{\mathcal{V}}_{\rm p}) \; ,
\label{complete momentum eq}
}
where $\vec{\xi}(\vec{r},t)$ is the mode displacement field, $H$ is the temperature scale height, $\delta S$ is the Lagrangian perturbation of specific entropy, $c_p$ is the specific heat capacity at constant pressure, $\rho$ is the equilibrium density, $\vec{\mathcal{V}}_{\rm p}(\vec{r},t)$ is the velocity field associated with the ensemble of plumes, \smash{$\vec{\mathcal{L}}^{\rm ad}$} is the adiabatic differential operator provided in \eq{L^ad}, \smash{$\vec{\mathcal{L}}^{\rm nad}$} is a differential operator given in \eq{L^nad} and resulting from non-adiabatic effects, $\partial_t$ denotes the partial time derivative, $\vec{\nabla}$ is the gradient operator, and ($\otimes$) is the outer product.

Monitoring the evolution of $\delta S$ requires to consider the perturbed heat equation. Accounting only for radiative losses in case of asymptotic gravity modes, this latter can be expressed within the diffusion approximation as (see \appendixname{}~\ref{diffusion})
\algn{
\partial_t \left( \frac{\delta S}{c_p} \right)=\frac{1}{t_{\rm R}} \left[ \mathcal{L}^{{\rm nad}1}\left( \frac{\vec{\xi}}{H}\right)+\mathcal{L}^{{\rm nad}2}\left( \frac{\delta S}{c_p}\right)\right] \; ,
\label{energy perturb eq 2}
}
where $t_{\rm R}= \rho c_p T H/F_{\rm R}$ is the local radiative thermal timescale, which describes the exchange rate of energy between the modes and the radiation, with  $F_{\rm R}$ and $T$ the equilibrium radiative flux and temperature, respectively. We note that \smash{$\mathcal{L}^{{\rm nad}1}$ and $\mathcal{L}^{{\rm nad}2}$} in the latter equation are dimensionless linear differential operators with respect to radius and represent the perturbation of the divergence of the radiative flux. We also emphasize that $t_{\rm R}$ is equal to the local thermal timescale in the radiative zone and in a very thin near-surface layer where the energy flux is mostly carried by radiation. This is not the case in deeper convective layers where the energy flux is mostly carried by the convective flux, but whose influence on the considered gravity modes can be neglected according to the numerical computations of \cite{Belkacem2009}.

\subsection{Mode amplitude in the quasi-adiabatic limit}

\subsubsection{Decomposition on the adiabatic eigenfunction basis}
\label{amplitude quasi-adiabatic}

The eigenfunctions $\vec{\xi}_{n\ell m}$ of the $\vec{\mathcal{L}}^{\rm ad}$ operator, with radial orders $n$, angular degrees $\ell$ and azimuthal numbers $m$, were shown to form a complete basis of the oscillation displacement field under the zero-boundary conditions \citep[e.g.,][]{Chandra1964,Unno1989}. More precisely, the demonstration relied on the assumption that the oscillations are adiabatic (i.e., $\delta S = 0$). For our purpose, we actually show in \appendixname{}~\ref{basis} that they also form a complete basis of the oscillation displacement field in the non-adiabatic case (i.e., $\delta S\ne 0$).
It is therefore possible to project the non-adiabatic mode displacement field onto the basis of the eigenfunctions of the $\vec{\mathcal{L}}^{\rm ad}$ operator. This gives
\algn{
\vec{\xi}(\vec{r},t)=\sum_{n=-\infty}^{+\infty}\sum_{\ell=0}^{+\infty} \sum_{m=-\ell}^{+\ell} a_{n\ell m}(t)~ \vec{\xi}_{n\ell m} (\vec{r}) \;,
\label{field expansion}
}
where we have introduced the instantaneous amplitude $a_{n\ell m}$.
The eigenfunctions satisfy the eigenvalue relation
\algn{
\vec{\mathcal{L}}^{\rm ad}\left( \vec{\xi}_{n\ell m}\right)=\omega_{n\ell m}^2 \vec{\xi}_{n\ell m} \; ,
\label{eigenvalue}
}
with
\algn{
\vec{\xi}_{n\ell m} (\vec{r})= \xi_{n\ell m}^r(r) ~Y_\ell^m(\theta,\varphi ) ~\vec{e}_r+\xi_{n\ell m}^h(r)~r\vec{\nabla} Y_l^m(\theta,\varphi )  \; ,
}
where $\omega_{n\ell m}$ is the angular eigenfrequency, $(r,\theta,\varphi)$ are the spherical coordinates in the stellar frame, $\xi_{n\ell m}^r$ and $\xi_{n\ell m}^h$ are the (real) radial and poloidal components of the eigenfunctions, respectively, which are normalized such as $\xi_{n\ell m}^r=1$ at the photosphere, and $Y_\ell^m$ are the orthonormal spherical harmonics. The eigenfunctions are orthogonal with respect to the density-weighted inner product and their mode mass is defined as
\algn{
\mathcal{M}_{n\ell m}\equiv \int_{V} \rho ~ \vec{\xi}_{n\ell m}\cdot \vec{\xi}_{n\ell m}^\star \dd V  \; ,
\label{orthogonality}
}
where $V$ is the stellar volume beyond which the stellar density vanishes and (${}^\star$) denotes the complex conjugate.

\subsubsection{Globally quasi-adiabatic gravity modes}
\label{globally quasi-adiabatic}

Asymptotic gravity modes are incompressible \citep[e.g.,][]{Dintrans2001}, so that \smash{$\mathcal{L}^{{\rm nad}1}$ and $\mathcal{L}^{{\rm nad}2}$} in \eq{energy perturb eq 2} are dominated by second-order derivatives with respect to radius and their local norms scale as $H^2/\lambda_{n\ell m}^2$ when applied on a harmonic $(n,\ell,m)$, where $\lambda_{n\ell m}(r)$ is the local radial wavelength (see \appendixname{}~\ref{scaling}). Consequently, the time evolution of the amplitude $a_{n\ell m}$ is simultaneously governed by
the local damping timescale given by \smash{$t_{\rm damp}\sim(\lambda_{n\ell m}^2/H)^2t_{\rm R}$} in \eq{energy perturb eq 2} and the dynamical timescale \smash{$t_{\rm dyn} \sim 1/\omega_{n\ell m}$} in \eq{complete momentum eq}.
Almost everywhere in the star, the quasi-adiabatic limit is supposed to be met, that is,
\algn{
t_{\rm dyn}  \ll  t_{\rm damp}(r) \; ,
}
and the non-adiabatic effects represented by the quantity $\delta S/c_p$ can be locally treated as a small perturbation \citep[e.g.,][]{Dziembowski1977b}. This is nevertheless not met in a very thin near-surface layer where the density vanishes and $t_{\rm R}$, which is equivalent to the thermal timescale in this region, becomes much smaller than dynamical timescale \citep[e.g., see Fig.~2b of][]{Berthomieu1990}. However, this region is so thin and the part of the total mode energy contained inside is so small that its impact on the global mode damping is expected to be negligible. This expectation is again supported by the numerical computations of \cite{Belkacem2009}, who demonstrated that the work performed by the radiative flux variations on the oscillations mainly originates from the radiative cavity for the considered frequency range. 

In order to reason in a global way, we therefore define a global damping timescale, denoted with $T_{\rm damp}$, which aims to measure the impact of the non-adiabatic effects on the global behavior of a harmonic $(n,\ell,m)$. Owing to the place of $t_{\rm damp}$ in \eq{energy perturb eq 2}, $T_{\rm damp}$ is taken equal to the inverse of the harmonic mean of $1/t_{\rm damp}(r)$ weighted by the local mode energy. Considering only the contribution from the radiative zone to the mode damping, $T_{\rm damp}$ reduces to \eq{T_damp lambda 2} in the asymptotic frequency range, that is,
\algn{
T_{\rm damp}^{-1} \approx \dfrac{\int_0^{r_{\rm b}} t_{\rm damp}^{-1} \dd r/\lambda_{n\ell m}}{\int_0^{r_{\rm b}} \dd r/\lambda_{n\ell m}} \; .
\label{T_damp lambda 1}
}
where $r_{\rm b}$ denotes the radius of the base of the convective zone. In the following, we thus assume that the oscillations are globally quasi-adiabatic in the considered frequency range, that is,
\algn{
t_{\rm dyn} \ll T_{\rm damp} \; .
}
The definition in \eq{T_damp lambda 1} will appear to be relevant and will ease the computation of the mode amplitude.

\subsubsection{Forced amplitude}
\label{forced amplitude}

Projecting \eq{complete momentum eq} on $\vec{\xi}_{n\ell m}$ and expressing $\delta S$ by means of \eq{energy perturb eq 2}, a set of coupled third-order linear differential equations for the dependent variables $a_{n\ell m}(t)$ can then be obtained, each of them accounting for the first-order non-adiabatic perturbations (see \appendixname{}~\ref{quasi-adiabatic}). These equations turn out to involve a fast timescale, which is $t_{\rm dyn}$, and a slow timescale, which is actually $T_{\rm damp}$ defined in \eq{T_damp lambda 1}. It thus appears judicious to solve the differential system using a two-timing method \citep[e.g.,][]{Kevorkian1961}. Such a method will provide us with a uniformly valid solution up to timescales of the order of $T_{\rm damp}$. A two-timing analysis is sufficient for the present purpose since most of the mode energy is dissipated on a timescale of the order of $T_{\rm damp}$, so that the error made is expected to be small compared to other uncertainties related, for instance, to the modeling of the generation process.

Expressing explicitly the partial derivatives with respect to the fast and slow timescales in the amplitude equations and grouping together the terms of the same order, a two-timing approach ultimately leads for the mode amplitude (see \appendixname{}~\ref{two-timing})
\algn{
a_{n\ell m}(t) \approx A_{n\ell m}(t)~e^{+i\omega_{n\ell m} t-\eta_{n\ell m} t}+ A_{n\ell m}^\star(t)~e^{-i\omega_{n\ell m} t-\eta_{n\ell m} t} \; ,
\label{a_n solution}
}
with
\algn{
A_{n\ell m}(t)\approx \frac{1}{2 i}\int_{-\infty}^t \omega_{n\ell m} \widetilde{\mathcal{F}}_{n\ell m}(t^\prime) e^{-i\omega_{n\ell m} t^\prime+\eta_{n\ell m} t^\prime} \dd t^\prime \; ,
\label{a_n forced}
}
where $\widetilde{\mathcal{F}}_{n\ell m}$ is provided by \eq{mode driving}. This latter term results from the scalar projection of the plume driving term in \eq{complete momentum eq} on $\vec{\xi}_{n\ell m}$ normalized by the mode mass. Besides, the expression of the damping rate $\eta_{n\ell m}$ is compatible, within the asymptotic limit, to the usual quasi-adiabatic expression provided, for example, by \cite{Dziembowski2001} or \cite{Godart2009}. Its expression is provided in \eqs{Godart 1}{Godart 2}~from which we deduce that \smash{$\eta_{n\ell m}\sim 1/T_{\rm damp}$}. This confirms a posteriori our choice for the definition of the global damping timescale in \eq{T_damp lambda 1}.

At this point, we note that \eq{a_n solution} can be retrieved if we assume that the solutions of the homogeneous equations for each amplitude $a_{n\ell m}$ take the form of an exponentially-damped harmonic oscillators, as done in usual quasi-adiabatic analyses \citep[e.g.,][]{Dziembowski1977b,Unno1989}, and then use this ansatz to find the particular solution of the forced equations. Nevertheless, the present analysis clearly highlights in addition two essential points. First, \eq{a_n solution} is uniformly valid up to a timescale of the order of $T_{\rm damp}$ only. Second, it appears that \eq{a_n solution} holds true if the coupling induced by the \smash{$\vec{\mathcal{L}}^{\rm nad}$} operator in \eq{complete momentum eq} between the different $(n,\ell,m)$ components remains negligible. Assuming that modes with adjacent radial orders $n$ and $n+1$ have comparable amplitudes for given values of $\ell$ and $m$, this is met if we have (see \appendixname{}~\ref{two-timing})
\algn{
\frac{\Delta t_{\rm core}}{T_{\rm damp}} \approx n \frac{t_{\rm dyn} }{ T_{\rm damp}}  \ll1 \; ,
\label{no coupling}
}
where $\Delta t_{\rm core}$ is the time spent by a wave energy ray of frequency $\omega_{n\ell m}$ to cross the radiative core. In this case, the radiative losses accumulated during a one-travel path of a wave through the radiative core are negligible; as a result, the resonant modes keep at leading order the same global structure as the eigenfunctions of \smash{$\vec{\mathcal{L}}^{\rm ad}$}, except for the small exponential decay of their amplitudes with time (see discussion in \appendixname{}~\ref{two-timing}). We will show in \sectionname{}~\ref{apparent} that this is met for a solar model in the considered frequency range. The relative error made when writing \eq{a_n solution} thus turns out to be of the order of \smash{$\Delta t_{\rm core}/T_{\rm damp}$} at most.

To express further \eq{a_n solution}, we then need to model the \smash{$\widetilde{\mathcal{F}}_{n\ell m}$} term and thus to specify the plume velocity field.

\subsection{Modeling the plumes and the penetration zone}
\label{plume model}

While the mean ensemble behavior of convective plumes was widely studied in numerical simulations -- for example, how they structure together, how they can merge or how efficiently they can transport energy \citep[e.g.,][]{Hardenberg2008,Pieri2016,Pratt2017} --, there is however no quantitative description of their detailed structures to the authors' knowledge. Moreover, though useful to guide the investigations, the outcomes of the current numerical simulations of extended convective envelopes still have to be taken with caution \cite[e.g.,][]{Rempel2004}.

In the present work, we thus choose to follow an analytical description based on the model of \cite{Pincon2016}. The penetration of convective plumes is supposed to be a random, stationary and ergodic process. We assume that, on average, $\mathcal{N}$ identical plumes (i.e., with the same velocity and shape) are penetrating at each time and can transfer their energy into gravity modes. The plumes are supposed to be incoherent with each other and uniformly-distributed on the sphere. At the base of the convective region, the Péclet number, which represents the ratio of the efficiency of the advection of heat by the plumes to the efficiency of the radiative diffusion, is expected to be much larger than unity in the Sun (see \sectionname{}~\ref{standard}). As a result, the convective plumes are braked over a very small penetration length (i.e., relatively to the local pressure scale height) below the radius where the Schwarzschild's criterion is met \citep{Zahn1991,Dintrans2005}. The plume ram pressure gradient is thus supposed to be maximum inside the penetration region, and to vanish outside. The analytical form of the velocity field inside the penetration region that is associated with a plume penetrating at $t=0$ and whose center has for latitudinal and azimuthal coordinates $(\theta_0 ,\varphi_0)$ reads \citep[][cf. Eqs.~(19)-(20)]{Pincon2016}
\begin{equation}
\vec{\mathcal{V}}_{{\rm p},0}(\vec{r},t)=f\left(\frac{t}{\tau_{\rm p}}\right)~\mathcal{V}_r( r)~ e^{-S_h^2/2b^2}~ \vec{e}_r \; ,  
\label{profil_Vp}
\end{equation} 
where the $f$ function represents the time evolution of the plume velocity, with $\tau_{\rm p}$ the plume lifetime, $\mathcal{V}_r$ is the radial profile, $b$ is the plume radius, and $S_h(\vec{r};\theta_0,\varphi_0)$ corresponds to the distance on a concentric sphere from the center of the plume.

Owing to the large Péclet number, an adiabatic temperature gradient is imposed in the penetration zone; $\mathcal{V}_r$ can thus follow the model of penetrative convection of \cite{Zahn1991}. The value of the plume velocity at the entry of the penetration region, denoted with $V_{\rm b}$, and that of the radius $b$ are adapted from the model of turbulent plumes developed by \cite{Rieutord1995}. Below, the transition from an adiabatic to a radiative temperature gradient is supposed to be very sharp, so that the Brunt-Väisälä frequency discontinuously changes from about zero in the adiabatic region to $N_{\rm t}$ at the top of the radiative zone. We emphasize that such an hypothesis is supported by the recent seismic inversions of the Brunt-Väisälä frequency in the Sun performed by \cite{Buldgen2020}. Indeed, their observations showed that the transition from an adiabatic to a radiative gradient at the base of the convective zone occurs over a distance $d_{\rm trans}$ representing about 0.5\% of the solar radius and that the value of $N_{\rm t}$ is equal to about $550~\mu$Hz. As a result, we find that $d_{\rm trans}$ is much smaller than the local wavelength in the considered ranges of frequencies and degrees. This justifies the fact that, from the point of view of the gravity modes, the profile of the Brunt-Väisälä frequency can be supposed to be discontinuous at the top of the radiative zone.

Besides, the time evolution of the plumes in the penetration region is badly understood. \cite{Pincon2016} identified two probable plume destruction processes: baroclinic instabilities or turbulence. Using orders of magnitude, they estimated that the plume lifetime is most likely to range around the convective turnover timescale of turbulent eddies at the base of the convective region, $\tau_{\rm conv}$. Regarding the profile, they assumed that the penetration of a plume is a very punctual event and follows a Gaussian law in time, as in the initial work of \cite{Townsend1966}. Nevertheless, given the related uncertainties, it is relevant to test other prescriptions. Within the framework of the stochastic excitation of gravity modes, the spectral density of turbulent kinetic energy of convective eddies was usually assumed to be Gaussian too until \cite{Belkacem2009}, based on 3D numerical simulations, showed that a Lorentzian form is more appropriate. Gaussian and Lorentzian laws represent two limiting cases in the time Fourier domain, since both stand for very rapidly and very slowly decreasing functions of frequency, respectively. By analogy, the spectral density of kinetic energy of the plumes will be also supposed to be Lorentzian, which, in the time domain, is equivalent to assume an exponential time profile. Therefore, the $f(t)$ function in \eq{profil_Vp} will be equal in both limiting cases to either
\algn{ f_{\rm G}\left(\frac{t}{\tau_{\rm p}}\right)\equiv e^{- t^2/\tau_p^2} ~~~~\mbox{or}~~~~f_{\rm E}\left(\frac{t}{\tau_{\rm p}}\right)\equiv e^{- |t|/\tau_p} \; .
\label{time_profile}
}
%

\subsection{Mean mode energy}
\label{asymptotic expression}

Owing to the random properties of the generation process, a statistical approach has to be considered. As the time-averaged mode amplitude vanishes, we need to estimate either the mean mode square amplitude or the mean mode energy.
We find more appropriate to reason in a first step on the mean mode energy since it does not depend on the chosen normalization of the eigenfunction basis whereas the amplitude does. 
Because of the incoherence of the convective plumes between each other, we first note that the wave velocity fields generated by different plumes cancel out with each other. Assuming in addition that the excitation is a stationary process and the plumes are all identical and uniformly distributed over the sphere, we show in \appendixname{}~\ref{general expression} that the total mean mode energy is merely equal to the oscillation energy generated by one single plume penetrating at $t=0$ and averaged over its angular position $(\theta_0,\varphi_0)$, multiplied by the instantaneous number of penetrating plumes $\mathcal{N}$.
Therefore, using \eq{profil_Vp} in \eq{a_n solution}, it is possible to compute the mean oscillation energy associated with each orthogonal $(n,\ell,m)$ harmonic. In the asymptotic limit, the WKB form of the eigenfunctions can be used to formulate analytically the result (see \appendixname{}~\ref{asymptotic form}).
In the case of a large Péclet number at the base of the convective region, that is, for a very small penetration region, we find that the mean energy of the $(n,\ell,m)$ harmonic can be finally expressed as (see \appendixname{}~\ref{simplified})
\begin{align}
\langle E_{n\ell m} \rangle &\approx \dfrac{  \left[(\omega_{n\ell m} \Delta \Pi_\ell/\pi^2)~\overline{L_{\rm p}}~ F_{{\rm d},\ell} ~e^{-\ell(\ell+1) b^2/2 r_{\rm b}^2} ~\mathcal{C}_{n\ell m}\right]}{2\eta_{n\ell m}} \; ,
\label{luminosity approximate}
\end{align} 
with
\algn{
\overline{L_{\rm p}}=\mathcal{A} \mathcal{S}_{\rm p} \frac{\rho_{\rm b} V_{\rm b}^3}{2} \; ,
\label{L_p mean}
}
where $\Delta \Pi_\ell$ is the asymptotic period spacing between two consecutive adiabatic gravity modes of degree $\ell$ given in \eq{quantization}, \smash{$\overline{L_{\rm p}}$} is the mean plume kinetic luminosity through the shell of radius $r_{\rm b}$ at the base of the convective zone, \smash{$\mathcal{A}=\mathcal{N} b^2/4r_{\rm b}^2$} is the plume filling factor, $\mathcal{S}_{\rm p}=\pi b^2$ is the area occupied by a single plume, $\rho_{\rm b}$ and $V_{\rm b}$ represent the density and the plume velocity at $r_{\rm b}$, and $F_{{\rm d},\ell}=V_{\rm b} k_{h,{\rm b}}/ N_{\rm t}$
is the Froude number at the top of the radiative zone, with \smash{$k_{h,{\rm b}}=\sqrt{\ell(\ell+1)} /r_{\rm b}$} the horizontal wavenumber of the mode.

We note that the numerator of \eq{luminosity approximate} represents the amount of power injected into the mode per unit of time. Inside, the term in brackets results from the mode mass. The Gaussian term represents the horizontal correlation between the plumes and the mode, while $\mathcal{C}_{n\ell m}$ measures the temporal correlation. The general analytical expression of $\mathcal{C}_{n\ell m}$ is provided by either \eq{C_G} in case of a Gaussian plume time evolution (i.e., $f=f_{\rm G}$) or \eq{C_E} in case of an exponential plume time evolution  (i.e., $f=f_{\rm E}$). In the considered frequency range, we usually have $\eta_{n\ell m} \ll \nu_{\rm p}\ll \omega_{n\ell m}$, where $\nu_{\rm p}=1/\tau_{\rm p}$. As a result, $\mathcal{C}_{n\ell m}$ is merely equal to (see \appendixname{}~\ref{temporal correlation})
\algn{
\mathcal{C}_{n\ell m} \approx 4\sqrt{\pi} ~\frac{\eta_{n\ell m}}{\nu_{\rm p}} ~\frac{\nu_{\rm p}^3 }{\omega_{n\ell m}^3}~~~~\mbox{if}~~~~f=f_{\rm G} \; ,
\label{C_G main}
}
and
\algn{
\mathcal{C}_{n\ell m} \approx 16 \frac{\nu_{\rm p}^3}{\omega_{n\ell m}^3}~~~~\mbox{if}~~~~f=f_{\rm E} \; .
\label{C_E main}
}
In the considered frequency range, the temporal correlation is thus expected to be much smaller in the Gaussian case than in the exponential case, so does the mean mode energy. This will be discussed in more details in \sectionname{}~\ref{apparent}.

\begin{figure}
\centering
\includegraphics[width=\hsize,trim= 0.5cm 0cm 0.8cm 1cm, clip]{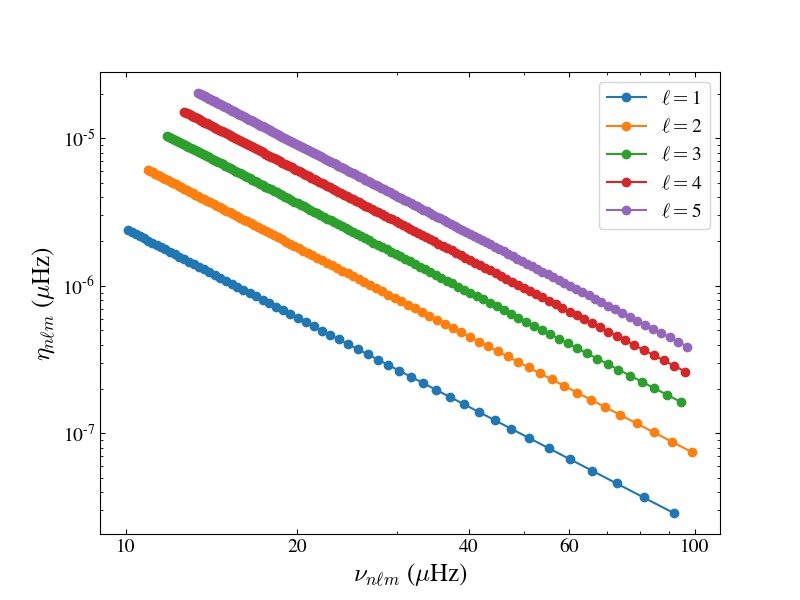} 
\caption{Radiative damping rate as a function of the oscillation frequency $\nu_{n\ell m}$ and the angular degree $\ell$. Each circle corresponds to a given eigenmode.}
\label{eta}
\end{figure}

Moreover, within the quasi-adiabatic and asymptotic limits, \smash{$\eta_{n\ell m}\propto \ell(\ell+1)/\omega_{n\ell m}^2$} as shown by \eq{Godart 2} and illustrated in \figurename{}~\ref{eta} \citep[e.g.,][]{Godart2009}.
In the considered frequency range, \eqss{luminosity approximate}{C_E main} thus demonstrate that $\langle E_{n\ell m}\rangle$ is independent of the frequency when $f=f_{\rm E}$ whereas $\langle E_{n\ell m}\rangle$ is inversely proportional to the squared frequency when $f=f_{\rm G}$. In both cases, the mean mode energy decreases as $\ell$ increases. As in the case of low-frequency progressive internal gravity waves, we see according to \eq{luminosity approximate} that the excitation efficiency is mainly measured by the Froude number at the top of the radiative region. This latter is expected to be much smaller than unity in stars (see \sectionname{}~\ref{input}); as the correlation and mode mass terms are also smaller than unity, the mean mode energy turns out to be much smaller than the mean plume kinetic energy at the base of the convective zone, and we check a posteriori that the feedback from the modes to the plume dynamics inside the penetration region remains negligible.

\subsection{Apparent mode radial velocity}
\label{rad vel}

Owing to a high signal-to-noise ratio, the search for the solar gravity modes is usually performed through radial velocity measurements integrated over the full solar disk \citep[e.g.,][]{Appourchaux2010}.
In order to be compared to the observed measurements, the theoretical mean mode energy has to be converted into mean mode radial velocity accounting for the line-of-sight projection and the limb-darkening effects \citep{Dziembowski1977,Berthomieu1990}. Among other observational effects that need to be corrected, these latter are expected to be predominant. Following the computation of \cite{Belkacem2009}, we show in \appendixname{}~\ref{mean velocity} that the mean apparent radial mode velocity for a $(n,\ell,m)$ harmonic is equal, in the slow rotator limit, to
\algn{
\varv_{n\ell m}^{\rm app}=\sqrt{\frac{\langle E_{n\ell m}\rangle}{\mathcal{M}_{n\ell m}}}  ~\tilde{V}_{n\ell m} \; ,
\label{v_app_2}
}
where $\tilde{V}_{n\ell m}$ is identified as
\algn{
\tilde{V}_{n\ell m}=\left|\alpha_\ell^m\xi_{n\ell m}^r(r_{\rm abs})+\beta_\ell^m \xi_{n\ell m}^h(r_{\rm abs}) \right| \; ,
\label{visibility}
}
with $r_{\rm abs}$ the radius in the atmosphere where the absorption line considered to measure the radial velocity is formed. The visibility coefficients $\alpha_\ell^m$ and $\beta_\ell^m$ are provided in \eqs{alpha_lm}{beta_lm}. They depend on the limb-darkening law as well as the angle between the stellar rotation axis and the line of sight.

We stress that, contrary to the mode energy, it is not possible to express further \eq{visibility} using the WKB form of the eigenfunctions as it becomes questionable in the convective region and the atmosphere of stars. The estimate of \eq{visibility} therefore requires the numerical computation of the mode eigenfunctions and their mode masses. 

%

\section{Applications to gravity modes in the Sun}
\label{apparent}

In this section, the excitation model is used to predict the apparent surface radial velocity of gravity modes generated by penetrative convection for a solar model. As in \cite{Belkacem2009}, we choose to compare our results to the detection threshold of the GOLF instrument on board of the SoHO spacecraft \citep[e.g.,][]{Gabriel1995}.

\subsection{Solar model}
\label{input}

We consider a calibrated solar model computed with the stellar evolution code CESTAM \citep{Marques2013}. The chemical composition follows the solar mixture such as given in \cite{AGS2009}, with initial helium and metals abundances $Y_0=0.25$ and $Z_0=0.013$. The OPAL 2005 equation of states and opacity tables as well as the NACRE nuclear reaction rates were used to build the model. The atmosphere was constructed following an Eddington gray approximation and convection was modeled using the mixing-length theory with a parameter $\alpha_{\rm MLT}=1.62$. Neither microscopic diffusion nor overshooting nor rotation were taken into account. We emphasize that the obtained solar structure, although rather approximative from the point of view of seismic inversions, is sufficient for our purpose as the uncertainties related to the assumptions considered in the excitation model are dominant. For example, it is known that microscopic diffusion can slightly modify the stratification at the base of the convective zone \citep[e.g.,][]{JCD1993} and hence can affect the properties of the penetrating plumes. Nevertheless, this effect is expected to be much smaller than the uncertainties related, for instance, to the plume formation mechanism across the convection zone.

The adiabatic eigenfrequencies $\nu_{n\ell m}=\omega_{n\ell m}/2\pi$, eigenfunctions and mode masses needed to compute \eq{v_app_2} were obtained via the oscillation code ADIPLS \citep[][]{JCD2008}.
The eigenfunctions were normalized such as \smash{$\xi_{n\ell m}^r= 1$} at the photosphere.
Owing to the large drop in the density and pressure at the stellar surface, the reflective zero-boundary conditions are considered. We check that the period spacing between consecutive $\ell=1$ eigenmodes is very close to the asymptotic value, that is, $\Delta\Pi_{\ell=1} \approx 26$~min. 
Moreover, the mode mass is about proportional to \smash{$\ell^{3}$} and \smash{$\nu_{n\ell m}^{-6}$} \citep[e.g., see Fig. 6 of][]{Belkacem2009}.
To be complete, the radiative damping rate occurring in \eq{luminosity approximate} and computed according to the quasi-adiabatic expression provided in \eq{Godart 2} is plotted in \figurename{}~\ref{eta}. In this figure, we first see that the damping rate is very similar to that computed numerically by \citet[][see Fig.~9]{Belkacem2009} who accounted in addition for the influence of the upper layers. This fact confirms our assumption that the work performed in the inner radiative cavity is the main contributor to the damping of the asymptotic gravity modes. Second, we also see in this figure that the damping timescale $T_{\rm damp} \sim 1/\eta_{n\ell m}$ is more than six orders of magnitude higher than the oscillation timescale $t_{\rm dyn}\sim 1/\omega_{n\ell m}$. These facts justify a posteriori the use of the global quasi-adiabatic approximation in this frequency range. Moreover, since $n\lesssim 10^3$ in the considered frequency range, it is obvious that $\Delta t_{\rm core} \approx n t_{\rm dyn} \ll T_{\rm damp}$, so that the hypothesis made in \eq{no coupling} appears a posteriori to be valid too (i.e., the coupling induced by non-adiabatic effects between adjacent radial orders is negligible up to a timescale of the order of $T_{\rm damp}$).

\subsection{GOLF apparent radial velocity with standard parameters}
\label{standard}

The internal structure of the solar model provides us with all the equilibrium quantities to compute the parameters of the excitation model and the mode apparent velocity in \eq{v_app_2}. The radius at the base of the convective zone and the local density are equal to $r_{\rm b}\approx 5.2~10^{5}$ km and $\rho_{\rm b} \approx 125$ kg~m$^{-3}$. Based on the model of plumes of \cite{Rieutord1995}, we find $b\approx10^4$ km and $V_{\rm b} \approx 190 ~\mbox{m~s}^{-1}$. The Péclet number at the base of the convective region is estimated to be on the order of $V_{\rm b} H(r_{\rm b}) / K_{\rm rad}(r_{\rm b}) \sim 10^7$, where $H$ and $K_{\rm rad}$ are the temperature scale height and the radiative diffusivity, respectively. The large Péclet number assumption used in \sectionname{}~\ref{plume model} is thus justified for the Sun. Following \cite{Buldgen2020}, the Brunt-Väisälä frequency just below $r_{\rm b}$ is taken equal to $N_{\rm t} = 550~\mu$Hz. The Froude number at the top of the radiative zone is thus equal to $F_{\rm {\rm d},1}\approx 10^{-3}$. Moreover, we use a reasonable value $\mathcal{N} \approx 1000$, as previously estimated by \cite{Rieutord1995}, which corresponds to $\mathcal{A} \approx 0.08$, in qualitative agreement with previous numerical simulations \citep[e.g.,][]{Stein1998,Brummell2002}. The convective turnover timescale at the base of the convective region is equal to about $\tau_{\rm conv}\approx10$ days according to the convection mixing-length theory, which leads to $\nu_{\rm p} \approx 1/\tau_{\rm conv} \approx1~\mu$Hz. At this point, the considered values of the plume parameters are referred to as the standard values. We consider that the sodium NaD1 and NaD2 absorption lines used by the GOLF instrument to measure radial velocities form at height $h\approx 300$km above the photosphere \citep{Bruls1992}. We also use the limb darkening law as formulated by \cite{Ulrich2000} and take for the angle between the stellar rotation axis and the line of sight $\Theta_0 \approx 83^\circ$. Moreover, we systematically focus on the azimuthal numbers $|m|=\ell$ in what follows as we find they are the less attenuated by the visibility effects and we want to study the more optimistic case.

\begin{figure*}
\centering
\includegraphics[scale=0.8,trim= 0cm 0cm 0cm 0cm, clip]{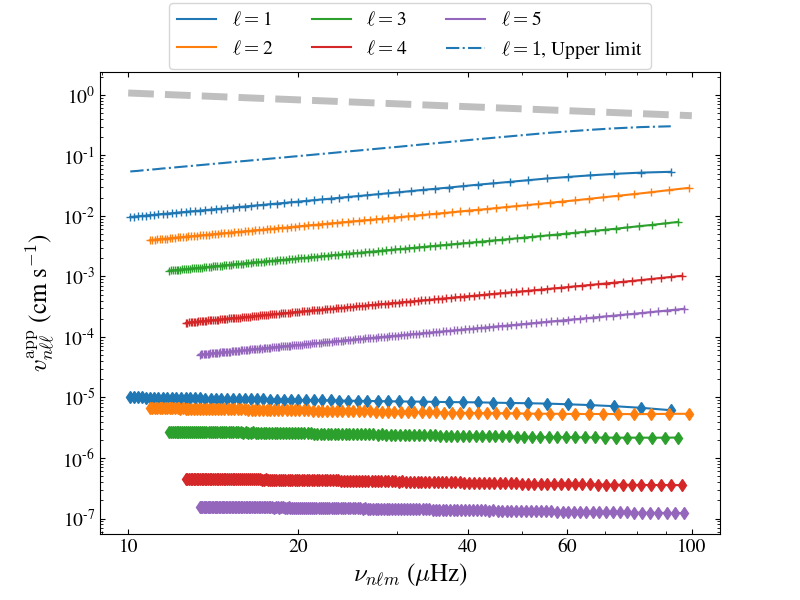} 
\caption{Apparent surface radial velocity of solar gravity modes as a function of the oscillation frequency $\nu_{n\ell m}$ and the angular degree $\ell$, for azimuthal numbers $m=\ell$ and for standard plume parameters. The results obtained with a Gaussian and an exponential plume time evolution (see \sectionname{}~\ref{plume model}) are represented by the diamonds and the plus symbols, respectively, for each considered eigenmode.
The thick dashed gray line corresponds to the 22-years GOLF detection threshold (see \sectionname{}~\ref{threshold}). The blue dash-dotted line represents the upper limit for the amplitude of the dipolar modes when accounting for the largest plausible variations in the plume parameters (i.e., with an increase in the inverse plume lifetime and radius by a factor of two, see \sectionname{}~\ref{sensitivity}).
}
\label{v_app_1}
\end{figure*}

Using all these physical ingredients, we can compute the apparent radial velocity of the asymptotic gravity modes, $\varv^{\rm app}_{n\ell\ell}$, between $10~\mu$Hz and $100~\mu$Hz. The result is potted in \figurename{}~\ref{v_app_1} as a function of the mode eigenfrequencies and for typical degrees $\ell=1-5$. Both limiting cases of a Gaussian and exponential plume time evolution are considered and compared.

First, \figurename{}~\ref{v_app_1} shows in both cases that the apparent surface velocity of the gravity modes generated by penetrative convection is maximum for $\ell=1$ and sharply drops as a function of $\ell$. Indeed, at a given frequency, the difference of apparent velocity between the $\ell=1$ and $\ell=5$ modes is larger than about two orders of magnitude. Besides, $\varv^{\rm app}_{n\ell\ell}$ appears to depend linearly on the frequency in the exponential case, while it is quasi independent of frequency in the Gaussian case. In order to disentangle the reasons for such trends from the different terms occurring in \eq{v_app_2}, we propose to rewrite \eq{visibility} in a more simple way. Owing to the reflective boundary conditions, the Lagrangian perturbation of pressure vanishes at the solar surface, which leads to \citep[e.g.][]{Unno1989}
\algn{
\xi_{n\ell m}^h(R_\odot) = \frac{GM_\odot}{R_\odot^3 \omega_{n\ell m}^2} \xi_{n\ell m}^r(R_\odot)\approx \left(\frac{100~\mu{\rm Hz}}{\nu_{n\ell m}}\right)^2 \; ,
\label{0bound}
}
where $G$ is the gravitation constant, $M_\odot$ and $R_\odot$ are the solar mass and radius, and where we considered the normalization \smash{$\xi_{n\ell m}^r(R_\odot)= 1$}.
Therefore, over the considered frequency range, we have \smash{$\xi_{n\ell m}^h(R_\odot)\ge \xi_{n\ell m}^r(R_\odot)$}. Moreover, we can assume that $\xi_{n\ell m}^h$ does not significantly vary in the solar atmosphere, that is, \smash{$\xi_{n\ell m}^h(r_{\rm abs}) \approx \xi_{n\ell m}^h(R_\odot)$}. We checked the validity of this approximation using the numerical eigenfunctions. Since \smash{$|\alpha_{\ell}^\ell|$} is about equal or smaller than $|\beta_{\ell}^\ell|$ (see \tablename{}~\ref{table1}), \eq{v_app_2} can thus be expressed at first approximation as
\algn{
\varv_{n\ell \ell}^{\rm app}\approx\sqrt{\frac{\langle E_{n\ell \ell}\rangle}{\mathcal{M}_{n\ell \ell}}}  ~\left|\beta_\ell^\ell\right| \left(\frac{100~\mu{\rm Hz}}{\nu_{n\ell \ell}~(\mu{\rm Hz})}\right)^2\; .
\label{v_app_3}
}
For $\ell=1-5$, the Gaussian term in \eq{luminosity approximate} remains close to unity since $\ell \ll r_{\rm b}/b\sim 50$. As a consequence, since $\eta_{n\ell m} \propto \ell(\ell+1)/\nu_{n\ell m}^2$, we find according to \sectionname{}~\ref{asymptotic expression} that \smash{$\langle E_{n\ell m} \rangle \propto \nu_{n\ell m}^{-2} $} in the Gaussian case and \smash{$\langle E_{n\ell m} \rangle \propto \ell^{-2} $} in the exponential case. In addition, as the mode mass depends approximately on $\nu_{n\ell m}$ and $\ell$ as \smash{$\mathcal{M}_{n\ell m} \propto \ell^{3} \nu_{n\ell m}^{-6}$}, \eq{v_app_3} leads in the Gaussian case to
\algn{
\varv_{n\ell \ell}^{\rm app}\propto \left|\beta_\ell^\ell\right| \ell^{-3/2}\; ,
\label{v_app_G}
}
and in the exponential case
\algn{
\varv_{n\ell \ell}^{\rm app}\propto \left|\beta_\ell^\ell\right| \ell^{-5/2} \nu_{n\ell \ell}\; .
\label{v_app_E}
}
As shown in \tablename{}~\ref{table1}, the values of $|\beta_\ell^\ell|$ smoothly varies with $\ell$, and the trends predicted in \eqs{v_app_G}{v_app_E}~are in qualitative agreement with \figurename{}~\ref{v_app_1}. It thus appears that the simultaneous influence of the mode mass and the damping rate counterbalances the decrease of the mode driving with frequency (via the temporal correlation term). This fact explains the trends observed in \figurename{}~\ref{v_app_1}. 

   \begin{table}[]
      \caption[]{Absolute values of the visibility factors of the GOLF instrument as a function of the angular degree $\ell$.}
         \label{table1}
     $$ 
         \begin{array}{cccccc}
           \hline
	  \hline
	\noalign{\smallskip}
            \ell& 1& 2&3&4&5\\
	\noalign{\smallskip}
            \hline
       \noalign{\smallskip}
	|\alpha_\ell^\ell|&0.19&0.12 &0.05&0.009&0.004\\
	\noalign{\smallskip}
	|\beta_\ell^\ell|&0.15&0.30 &0.25&0.07&0.04\\
	\noalign{\smallskip}
	\hline
         \end{array}
     $$ 
   \end{table}
%

Moreover, we clearly show in \figurename{}~\ref{v_app_1} that the value of $\varv^{\rm app}_{n\ell\ell}$ predicted assuming a Gaussian plume time evolution at a given $\ell$ is about three orders of magnitude at least smaller than that  assuming an exponential plume time evolution. This huge difference results from the much smaller temporal correlation between the plumes and the considered eigenmodes in case of a Gaussian time evolution, which is represented by the $\mathcal{C}_{n\ell m}$ term. Indeed, in this case, the spectral density of the plume kinetic energy is mostly carried by the frequencies lower than $\nu_{\rm p}$. As $\nu_{n\ell m} \gg \nu_{\rm p}$, the coupling between the plumes and the eigenmodes is very small and the excitation is very inefficient. In contrast, in case of an exponential time evolution, the spectral density of the plume kinetic energy is distributed over higher frequencies and the coupling between the plumes and the eigenmodes is maximum for frequencies around $\nu_{n\ell m}$  (see the computation in \appendixname{}~\ref{temporal correlation}). As a result, the energy transfer from the plumes to the modes is much more efficient. Considering \eqs{C_G main}{C_E main}~in \eqs{luminosity approximate}{v_app_2}, the ratio of the mode apparent velocity in the Gaussian case to that in the exponential case scales as \smash{$\sqrt{\eta_{n\ell m}/\nu_{\rm p}} \ll 1$}. Since $\eta_{n\ell m}\propto 1/\nu_{n\ell m}^{2}$ in the quasi-adiabatic limit, this ratio is expected to decrease as $1/\nu_{n\ell m}$, in agreement with \figurename{}~\ref{v_app_1}. Our estimate therefore demonstrates that the amplitude of the asymptotic solar gravity modes generated by penetrative convection critically depends on the plume time evolution at the base of the convective zone. Our lack of knowledge on the plume dynamics in this region thus prevents us from quantitatively predicting the efficiency of this excitation process. In turn, if detected, solar gravity modes can be expected to put important constraints on penetrative convection at the top of the radiative zone (see discussion in \sectionname{}~\ref{potential}).

Finally, we point out that over the considered frequency range, the GOLF apparent radial velocity of the gravity modes computed with the present model and standard plume parameters is about one order of magnitude smaller than that predicted considering turbulent pressure as the driving mechanism. Indeed, for the $\ell=1$ modes in the exponential case, for which the excitation is the most efficient, our predictions lie between $\varv^{\rm app}_{n\ell\ell}\approx 0.01-0.05$~cm~s${}^{-1}$ from $\nu_{n\ell m}=10~\mu$Hz to $\nu_{n\ell m}=100~\mu$Hz, while those of \cite{Belkacem2009} lie between $\varv^{\rm app}_{n\ell\ell}\approx 0.1-0.5$~cm~s${}^{-1}$  (see \sectionname{}~\ref{comparison} for a more detailed comparison). These results therefore suggest that penetrative convection is less efficient than turbulent convection to generate asymptotic gravity modes in the Sun. Nevertheless, owing to a significant sensitivity of the predictions to the plume parameters (i.e., velocity, radius, lifetime), the values of which currently suffer uncertainties, we emphasize that reasonable variations in these parameters are likely to reduce the gap between both excitation mechanisms (e.g., see blue dash-dotted line in \figurename{}~\ref{v_app_1}). This point is further discussed in \sectionname{}~\ref{sensitivity}.

%

\subsection{Comparison with the GOLF detection threshold}
\label{threshold}

Using a statistical approach, \cite{Appourchaux2010} could analytically express the threshold signal-to-noise ratio above which a peak in the power spectral density (PSD) of an observed time series can be considered as a relevant oscillating signal over a frequency interval $\Delta f$, with a false alarm probability $p_{\rm th}$ that the measurement is due to pure noise. As shown in \appendixname{}~\ref{threshold_app}, this threshold can be converted into a detection threshold for the root mean square oscillation velocity, denoted with $\varv_{\rm th}$, above which a signal would be detected with a certain confidence level. It reads
\algn{
\varv_{\rm th} \approx \sqrt{\ln\left(\frac{T  \Delta f}{ p_{\rm th} }\right)\frac{\tilde{s}}{T}}\; ,
\label{v_th}
}
where $T$ is the observation time and $\tilde{s}$ is the mean noise level in the PSD over the range $\Delta f$.

We consider a GOLF observation time equal to $T=22$ yrs. The mean noise level $\tilde{s}$ is estimated from \citet[][see \appendixname{}~\ref{threshold_app}]{Garcia2007}. We choose $\Delta f=10~\mu$Hz and $p_{\rm th}=0.01$. Instead of a common rejection level of 10\%, we adopt a rejection level of 1\% since it allows the posterior probability that a peak is due to noise to reach lower values close to 10\%, as already mentioned in \cite{Appourchaux2010}.
The resulting 22-years GOLF detection threshold is represented by a thick gray dashed line in \figurename{}~\ref{v_app_1}. We see that the threshold velocity smoothly decreases from about $1$~cm~s${}^{-1}$ at $\nu_{n\ell m} = 10~\mu$Hz to about $0.5$~cm~s${}^{-1}$ at $\nu_{n\ell m} = 100~\mu$Hz. This trend results from the decrease in the solar granulation noise with frequency in the GOLF data. Its value is about five orders of magnitude higher than the apparent mode velocity throughout the considered frequency range when assuming a Gaussian plume time evolution. According to \eq{v_th}, the amplitude of gravity modes will remain lower even if the observation time is equal to ten billion years. In the case of an exponential plume time evolution, the amplitude predicted with standard plume parameters remains also one order of magnitude at least lower the 22-years GOLF detection threshold over the considered frequency range. Based on our simple estimate of the GOLF mean noise level (see \eq{mean noise}), \eq{v_th} requires an unreachable GOLF observation time $T\sim 2000$~yrs to detect the dipolar $n=6$ gravity mode at $\nu_{n\ell m} \approx 90~\mu$ Hz. In conclusion, using standard values for the plume parameters, our model indicates that the solar asymptotic gravity modes generated by penetrative convection remains far from being detected in an acceptable observation time by radial velocity measurements such as performed by the GOLF instrument. (Correction) It is worth mentioning that the uncertainties on the predicted amplitude, which are inherent to uncertainties on the plume parameters, may nevertheless considerably reduce the observation time required for a reliable detection. This potential is also evaluated in \sectionname{}~\ref{sensitivity}.

\section{Discussion} 
\label{discussion}

\subsection{Comparison with the estimate of \cite{Andersen1996}}

\cite{Andersen1996} first investigated the generation of solar gravity modes by penetrative convection using both numerical simulations and simple energy considerations. Including an ad-hoc forcing term at the top of the radiative region that mimics the influence of penetrative plumes in a solar model, he numerically solved the wave equations and computed the transmission of the generated wave energy throughout the convective envelope, the structure of which was taken from the turbulent numerical simulations of \cite{Andersen1994}. He found that the wave transmission up to the surface is of the order of $0.05\%$. Using these results, he could estimate that the horizontal velocity of gravity modes near the solar surface ranges between $0.01-1~\rm{mm}~\rm{s}^{-1}$ for $\ell = 6$ and $\nu_{n\ell m}=50-100~\mu$Hz. When accounting for an appropriate GOLF visibility factor, that is $|\beta_6^6|\approx 0.03 $, this upper value turns out to be much larger than our predictions in case of a Gaussian plume time evolution, but of the same order of magnitude in the exponential case. However, we emphasize that this qualitative agreement is more likely to be fortuitous. Indeed, in order to estimate the amplitude of gravity modes, \cite{Andersen1996} assumed that the total mode energy is equal to the part of the convective kinetic energy transferred on average to progressive gravity waves in the numerical simulation of \cite{Andersen1994}, the upper and lower boundary conditions of which are open. As a consequence, this estimate actually did not properly account for the quasi stationarity of the modes, as well as the damping processes. Therefore, although qualitatively comparable for the $\ell=6$ gravity modes, the physical origins of our predictions radically differ from those of \cite{Andersen1996}, and drawing conclusions from their comparison ultimately appears irrelevant.

\subsection{Comparison with turbulence-induced gravity modes \citep{Belkacem2009}}
\label{comparison}

\cite{Belkacem2009} modeled the generation of asymptotic gravity modes by considering the turbulent Reynolds stress as the driving mechanism. As mentioned at the end of \sectionname{}~\ref{standard}, the surface mode velocity they predicted is one order of magnitude larger than our estimate in the exponential case when considering standard values for the plume parameters based on semi-analytical models and orders of magnitude. Nevertheless, for both excitation mechanisms, the dependence of the apparent radial velocities on $\ell$ and $\nu_{n\ell m}$ turns out to be comparable \citep[e.g.,][see Fig. 11]{Belkacem2009}.
This similar dependence results on the one hand from the mode mass and the damping rate, which play the same role in both excitation processes, and, on the other hand, from a comparable temporal correlation between the driving source and the modes. Indeed, \cite{Belkacem2009} assumed that the time coherence of the convective eddies is exponential; as a result, the temporal correlation between the modes and the eddies has a similar form to that between the modes and the plumes when assuming an exponential plume time profile \citep[e.g., a dependence on about $\nu_{n\ell m}^{-3}$, see Eqs.~(1)-(2) and Fig.~3 of][]{Belkacem2009}.
Regarding the magnitude, as the velocity of the turbulent eddies that drive gravity modes is about equal to $\varv_{\rm MLT} \approx60~$m~s${}^{-1}$ (according to the mixing length theory), which is about three times lower than the plume velocity at the base of the convective zone, penetrative convection could have been at first sight expected to generate gravity modes with higher amplitudes. However, while the excitation by penetrative convection occurs in a very thin shell at the base of the convective zone, convective eddies can drive gravity modes in a larger volume of the envelope. Moreover, as the eddy turnover frequencies are higher than $\nu_{\rm p}$ in upper layers of the convective region, the time correlation is larger for the excitation by turbulent convection than by penetrative plumes. The lower velocities of the convective eddies are thus compensated by a more extended excitation region and a better temporal correlation with the modes, whence higher amplitudes in the exponential case.

In contrast, when assuming that the eddy-time coherence is Gaussian as in \cite{Kumar1996}, \cite{Belkacem2009} found the apparent radial velocities of gravity modes generated by turbulent eddies is of the order of $10^{-3}~$cm~s${}^{-1}$. Although their predictions also depend on $\nu_{n\ell m}$ and $\ell$ in a similar way to our predictions in the Gaussian case, their magnitude is much larger than ours. Indeed, while the same arguments as in the previous paragraph hold true to explain the similarities in the dependence on $\nu_{n\ell m}$ and $\ell$, the temporal correlation with the modes in the Gaussian case is very sensitive to the timescales associated with the driving source. As the values of the turnover frequencies of the turbulent eddies driving the modes are larger than $\nu_{\rm p}$, the temporal correlation is expected to be much larger than for the excitation by plumes. This certainly explains the huge difference between the excitation by turbulent convection and penetrative convection in the Gaussian case.

\subsection{Sensitivity of the mode amplitude to the plume parameters}
\label{sensitivity}

As shown by \cite{Pincon2016}, the efficiency of the wave excitation by penetrative convection depends in a significant way on the plume parameters $b$, $\nu_{\rm p}$ and $V_{\rm b}$, whose values are subject to uncertainties.
For example, the plume model of \cite{Rieutord1995} is expected to provide an upper limit on $V_{\rm b}$ since it does not take into account the upward counterflow that can exchange momentum with the downward plumes and hence slow them down \citep[e.g.,][]{Rempel2004}. In contrast, the plume radius $b$ is likely to be underestimated since the model of \cite{Rieutord1995} does not account for the possible clustering of plumes, as observed for instance in numerical simulations \citep{Hardenberg2008}. In addition, as already discussed in \cite{Pincon2016}, turbulence inside the penetration region could lead to values of $\nu_{\rm p}$ substantially larger than the convective eddy turnover frequency such as predicted by the MLT.

To investigate the influence of these uncertainties on the mode amplitudes, we arbitrarily consider in this section the effect of a decrease in $V_{\rm b}$ of 30\%, and an increase in $\nu_{\rm p}$ and $b$ by a factor of two, while keeping the filling factor $\mathcal{A}$ constant (i.e., $\mathcal{N}\approx 250$ when the value of $b$ is twice larger).
Neglecting the variations of the horizontal correlation term in \eq{luminosity approximate} for $\ell=1-5$, \eqss{luminosity approximate}{v_app_2}~show that, at given frequency and degree, \smash{$\varv^{\rm app}_{n\ell\ell}\propto b V_{\rm b}^2 \nu_{\rm p}^{3/2}$}  in the exponential case and \smash{$\varv^{\rm app}_{n\ell\ell}\propto b V_{\rm b}^2 \nu_{\rm p}$} in the Gaussian case.
We see that a decrease in $V_{\rm b}$ of 30\% leads to a decrease by a factor of two in $\varv^{\rm app}_{n\ell\ell}$, while an increase of $b$ or $\nu_{\rm p}$ by a factor of two results in an increase by a factor of between two or three. As the mode amplitudes are insignificantly low in the Gaussian case, such variations in the parameters, and even much larger variations, do not modify the large gap with the exponential case and the GOLF detection threshold; gravity modes thus remain undetectable in this case. In the exponential case, such variations in the parameters $b$ and $\nu_{\rm p}$ can in contrast affect the predictions about the detectability of the modes. Indeed, assuming an increase in $\nu_{\rm p}$ by a factor of two, which results in an increase by a factor of about three in the amplitudes, the GOLF observation time required to detect a plume-induced dipolar gravity modes at $\nu_{n\ell m}=100~\mu$Hz is decreased by about one order of magnitude (i.e., to $T\sim 200~$yrs). Considering the largest plausible variations with a simultaneous increase in $b$ by a factor of two, the observation time required to detect such a mode is reduced to $T\sim 50~$yrs. In this most favorable case considered here, the predicted amplitudes of the dipolar plume-induced gravity modes are plotted in \figurename{}~\ref{v_app_1} (blue dash-dotted line). This upper limit turns out to be close to the amplitudes of the turbulent-induced gravity modes such as predicted by \cite{Belkacem2009}.

In conclusions, the time evolution of the plumes at the base of the convective zone represents the major source of uncertainties in our model. Indeed, the error related to the uncertainties on the plumes parameters is insignificant compared to the effect of the assumption on the plume time evolution. Besides, while the plume-induced gravity modes remain undetectable in the Gaussian assumption whatever the values of the plume parameters, the uncertainties on these parameters can significantly modify the predictions of the detectability of few modes in the exponential assumption. In the most favorable plausible case, the GOLF observation time required to detect the plume-induced $\ell=1$ gravity modes around $\nu_{n\ell m} \approx 100~\mu$Hz is reduced to $T\sim 50~$yrs, with amplitudes close to those predicted considering turbulent pressure as the driving mechanism.

\subsection{Can the amplitudes of gravity modes bring constraints on penetrative convection?}
\label{potential}

We see in \sectionname{}~\ref{sensitivity} that our lack of knowledge on penetrative convection affects our predictions. In turn, we can wonder to what extent the measurement of the amplitudes of gravity modes in the Sun can provide us with constraints on penetrative convection and the properties at the base of the convective zone.

Actually, our model shows either that the contribution from penetrative convection to the mode amplitude is undetectable for a Gaussian-like plume time evolution, or that it behaves similarly as a function of the frequency and the degree to the contribution from turbulent convection for an exponential-like plume time profile. As both excitation mechanisms are subject to uncertainties \citep{Belkacem2009}, it can thus appears difficult at first sight to disentangle both contributions and interpret the observed amplitudes in terms of structure without any independent information. Assuming that the plume time evolution is exponential-like, the measurement of the mode amplitude can nevertheless easily put upper limits on each excitation process, by ensuring that the predictions in each case remain lower than the observations. In fact, as the solar gravity modes have not been detected yet, we can similarly proceed by ensuring that the predictions remain lower than the current detection threshold. Indeed, imposing that $\varv_{\rm th} \gtrsim \varv^{\rm app}_{n\ell\ell}$ for $\nu_{n\ell m} = 100~\mu$Hz and $\ell=1$ in the exponential case, we find, using \eqs{luminosity approximate}{C_E main}{v_app_2}{v_th}~that for 22-years GOLF observations,
\algn{
F_{{\rm d},1}~\left(\frac{\overline{L_{\rm p}}}{L_\odot}\right) ~\left(\frac{\nu_{\rm p}}{1~\mu{\rm Hz}}\right)^{3}\lesssim 10^{-6} \; ,
}
where $L_\odot$ is the solar luminosity. 

Although the potential of the gravity mode amplitudes to probe penetrative convection is highlighted here, the present work suggests that the current theoretical uncertainties on the modeling of penetrative convection, and in particular on the plume time evolution, would limit their physical interpretation. The issue of disentangling the contributions from different sources to the excitation is in particular brought to the fore. Theoretical efforts, by means of numerical simulations and semi-analytical models, as well as possible complementary observational constraints, are needed in the future to improve our modeling of this phenomenon and elaborate relevant and robust seismic diagnoses to interpret the amplitude of gravity modes. 

\section{Conclusions} 
\label{conclusion}

In this work, we aim to estimate the amplitude of the asymptotic solar gravity modes generated by penetrative convection in the frequency range between $10~\mu$Hz and $100~\mu$Hz. Following \cite{Pincon2016}, we consider the ram pressure of an ensemble of incoherent, uniformly-distributed convective plumes penetrating into the top layers of the radiative zone as the driving mechanism. The forced oscillation equation is solved in the global quasi-adiabatic approximation using a two-timing method. As a result, we obtain an analytical expression of the mean mode energy, which is converted into apparent radial mode velocity through appropriate visibility factors.
The standard plume modeling (i.e., plume radius, velocity and lifetime) follows semi-analytical models and their time evolution in the penetration region is assumed to be either Gaussian or exponential, by analogy with stochastic excitation by turbulent eddies. The apparent mode radial velocity is computed for a solar model in both cases and the result is compared to the 22-years GOLF detection threshold.

We find that the mean mode energy drastically depends on the assumption about the time evolution of the plumes inside the penetration region. On the one hand, in the limiting case of a Gaussian time evolution, asymptotic gravity modes turn out to be undetectable by means of radial velocity measurements such as performed by the GOLF instrument. This is the consequence of a too large plume lifetime compared to the oscillation period. This result holds true despite a wide range of values considered for the parameters of the model. On the other hand, in the other limiting case of an exponential time evolution, we find that penetrative convection can generate gravity modes in a much more efficient way than in the Gaussian case. In this case, the lower the angular degree or the higher the frequency, the larger the apparent mode radial velocity. Using standard values for the plume parameters, the apparent radial mode velocity appears to reach about \smash{$0.05$~cm~s${}^{-1}$} for $\ell=1$ and $\nu_{n\ell m}\approx 100~\mu$Hz. These predictions are one order of magnitude smaller than those predicted considering turbulent pressure as the driving mechanism and remain well below the current 22-years GOLF detection threshold. Nevertheless, accounting for uncertainties in the plume parameters, we find in the most favorable plausible case that the predicted apparent mode radial velocity can be increased by a factor of six, that is, lying around \smash{$0.3$~cm~s${}^{-1}$} for $\ell=1$ and $\nu_{n\ell m}\approx 100~\mu$Hz. The observation time required to detect such a mode is reduced to about $50~$yrs with the GOLF instrument. These variations mainly result both from an important sensitivity of the mode amplitude to the plume parameters, and, in contrast, from a small sensitivity of the detection threshold to the observation time. Our findings thus indicate that, in the most favorable plausible case, penetrative convection can drive asymptotic gravity modes as efficiently as turbulent convection and with amplitudes close to the detection limit. We highlight that, if detected, the measurement of the gravity modes amplitude is expected to bring constraints on penetrative convection, but that the current uncertainties on the modeling of penetrative convection, and in particular their temporal evolution, will certainly limit their physical interpretation.
 
The results of this work call for further studies, either observational or theoretical. First, our estimates as well as previous ones about the excitation by turbulent pressure clearly suggest that we are likely to be very close to the detection in the asymptotic frequency range, and encourage carrying on efforts in observations and data analyses. While we mainly focused on the 22-years GOLF data, we note that a myriad of other data is available too and form an important source of information to be analyzed, as for instance the observations by the GONG and BiSON ground-based telescope networks. Second, it will be interesting in the future to extend the theoretical predictions to a higher frequency range. Indeed, a simple extrapolation of the available predictions toward slightly higher frequencies suggest that the amplitudes of the gravity modes in this domain are also likely to be close to the current detection limit. As already pointed out by \cite{Belkacem2011b}, predicting the amplitude of such low radial orders gravity modes will require to account consistently for the interplay between oscillations and convection, which is challenging since it will demand to combine a proper non-local time-dependent treatment of convection with a fully non-adiabatic treatment of pulsations. Furthermore, new developments, based both on numerical simulations and semi-analytical models, are needed to improve our understanding about the behavior of downward convective plumes at the interface with the radiative region. Though challenging as simulations remain far from the stellar regimes, the promising combination of such theoretical advancements with future measurements of the gravity mode amplitudes is hoped, for instance, to bring constraints on the dynamics and the mixing at work at the base of the convective zone.

\begin{acknowledgements}
We thank the anonymous referee for his careful reading and relevant comments that greatly helped improving the manuscript. We sincerely acknowledge K. Belkacem and M.-A. Dupret for the very interesting discussions on the present subject and their sensible comments. During this work, C. P. was funded by a postdoctoral fellowship of Chargé de Recherche from F.R.S.-FNRS (Belgium). G.B. acknowledges fundings from the SNF AMBIZIONE grant No 185805 (Seismic inversions and modelling of transport processes in stars). C. P. warmly thanks M. Huet, V. Huet, A. Leguillon and C. Houdmond for their sincere friendship and their encouragements during the writing of this paper.
\end{acknowledgements}

\bibliographystyle{aa} 
\bibliography{bib}

\begin{thebibliography}{73}
\expandafter\ifx\csname natexlab\endcsname\relax\def\natexlab#1{#1}\fi

\bibitem[{{Alvan} {et~al.}(2014){Alvan}, {Brun}, \& {Mathis}}]{Alvan2014}
{Alvan}, L., {Brun}, A.~S., \& {Mathis}, S. 2014, \aap, 565, A42

\bibitem[{{Andersen}(1994)}]{Andersen1994}
{Andersen}, B.~N. 1994, \solphys, 152, 241

\bibitem[{{Andersen}(1996)}]{Andersen1996}
{Andersen}, B.~N. 1996, \aap, 312, 610

\bibitem[{{Appourchaux} {et~al.}(2010){Appourchaux}, {Belkacem}, {Broomhall},
  {Chaplin}, {Gough}, {Houdek}, {Provost}, {Baudin}, {Boumier}, {Elsworth},
  {Garc{\'{\i}}a}, {Andersen}, {Finsterle}, {Fr{\"o}hlich}, {Gabriel}, {Grec},
  {Jim{\'e}nez}, {Kosovichev}, {Sekii}, {Toutain}, \&
  {Turck-Chi{\`e}ze}}]{Appourchaux2010}
{Appourchaux}, T., {Belkacem}, K., {Broomhall}, A.-M., {et~al.} 2010, \aapr,
  18, 197

\bibitem[{{Appourchaux} \& {Corbard}(2019)}]{Appourchaux2019}
{Appourchaux}, T. \& {Corbard}, T. 2019, \aap, 624, A106

\bibitem[{{Appourchaux} {et~al.}(2000){Appourchaux}, {Fr{\"o}hlich},
  {Andersen}, {Berthomieu}, {Chaplin}, {Elsworth}, {Finsterle}, {Gough},
  {Hoeksema}, {Isaak}, {Kosovichev}, {Provost}, {Scherrer}, {Sekii}, \&
  {Toutain}}]{Appourchaux2000}
{Appourchaux}, T., {Fr{\"o}hlich}, C., {Andersen}, B., {et~al.} 2000, \apj,
  538, 401

\bibitem[{{Appourchaux} \& {Pall{\'e}}(2013)}]{Appourchaux2013}
{Appourchaux}, T. \& {Pall{\'e}}, P.~L. 2013, Astronomical Society of the
  Pacific Conference Series, Vol. 478, {The History of the g-mode Quest}, ed.
  K.~{Jain}, S.~C. {Tripathy}, F.~{Hill}, J.~W. {Leibacher}, \& A.~A.
  {Pevtsov}, 125

\bibitem[{{Asplund} {et~al.}(2009){Asplund}, {Grevesse}, {Sauval}, \&
  {Scott}}]{AGS2009}
{Asplund}, M., {Grevesse}, N., {Sauval}, A.~J., \& {Scott}, P. 2009, \araa, 47,
  481

\bibitem[{{Basu} \& {Antia}(2008)}]{Basu2008}
{Basu}, S. \& {Antia}, H.~M. 2008, \physrep, 457, 217

\bibitem[{{Belkacem}(2011)}]{Belkacem2011b}
{Belkacem}, K. 2011, {Amplitudes of Solar Gravity Modes}, ed. J.-P. {Rozelot}
  \& C.~{Neiner}, Vol. 832, 139

\bibitem[{{Belkacem} {et~al.}(2009){Belkacem}, {Samadi}, {Goupil}, {Dupret},
  {Brun}, \& {Baudin}}]{Belkacem2009}
{Belkacem}, K., {Samadi}, R., {Goupil}, M.~J., {et~al.} 2009, \aap, 494, 191

\bibitem[{{Berthomieu} \& {Provost}(1990)}]{Berthomieu1990}
{Berthomieu}, G. \& {Provost}, J. 1990, \aap, 227, 563

\bibitem[{{B{\"o}ning} {et~al.}(2019){B{\"o}ning}, {Hu}, \&
  {Gizon}}]{Boning2019}
{B{\"o}ning}, V. G.~A., {Hu}, H., \& {Gizon}, L. 2019, \aap, 629, A26

\bibitem[{{Bonventre} \& {Orebi Gann}(2018)}]{Bonventre2018}
{Bonventre}, R. \& {Orebi Gann}, G.~D. 2018, European Physical Journal C, 78,
  435

\bibitem[{{Brookes} {et~al.}(1976){Brookes}, {Isaak}, \& {van der
  Raay}}]{Brookes1976}
{Brookes}, J.~R., {Isaak}, G.~R., \& {van der Raay}, H.~B. 1976, \nat, 259, 92

\bibitem[{{Bruls} \& {Rutten}(1992)}]{Bruls1992}
{Bruls}, J.~H.~M.~J. \& {Rutten}, R.~J. 1992, \aap, 265, 257

\bibitem[{{Brummell} {et~al.}(2002){Brummell}, {Clune}, \&
  {Toomre}}]{Brummell2002}
{Brummell}, N.~H., {Clune}, T.~L., \& {Toomre}, J. 2002, \apj, 570, 825

\bibitem[{{Buldgen}(2019)}]{Buldgen2019}
{Buldgen}, G. 2019, Bulletin de la Societe Royale des Sciences de Liege, 88, 50

\bibitem[{{Buldgen} {et~al.}(2020){Buldgen}, {Eggenberger}, {Baturin},
  {Corbard}, {Christensen-Dalsgaard}, {Salmon}, {Noels}, {Oreshina}, \&
  {Scuflaire}}]{Buldgen2020}
{Buldgen}, G., {Eggenberger}, P., {Baturin}, V.~A., {et~al.} 2020, \aap, 642,
  A36

\bibitem[{{Chandrasekhar}(1964)}]{Chandra1964}
{Chandrasekhar}, S. 1964, \apj, 139, 664

\bibitem[{{Christensen-Dalsgaard}(2008)}]{JCD2008}
{Christensen-Dalsgaard}, J. 2008, \apss, 316, 113

\bibitem[{{Christensen-Dalsgaard}(2020)}]{JCD2020}
{Christensen-Dalsgaard}, J. 2020, arXiv e-prints, arXiv:2007.06488

\bibitem[{{Christensen-Dalsgaard} {et~al.}(1993){Christensen-Dalsgaard},
  {Proffitt}, \& {Thompson}}]{JCD1993}
{Christensen-Dalsgaard}, J., {Proffitt}, C.~R., \& {Thompson}, M.~J. 1993,
  \apjl, 403, L75

\bibitem[{{Delache} \& {Scherrer}(1983)}]{Delache1983}
{Delache}, P. \& {Scherrer}, P.~H. 1983, \nat, 306, 651

\bibitem[{{Dintrans} {et~al.}(2005){Dintrans}, {Brandenburg}, {Nordlund}, \&
  {Stein}}]{Dintrans2005}
{Dintrans}, B., {Brandenburg}, A., {Nordlund}, {\AA}., \& {Stein}, R.~F. 2005,
  \aap, 438, 365

\bibitem[{{Dintrans} \& {Rieutord}(2001)}]{Dintrans2001}
{Dintrans}, B. \& {Rieutord}, M. 2001, \mnras, 324, 635

\bibitem[{{Dziembowski}(1977{\natexlab{a}})}]{Dziembowski1977}
{Dziembowski}, W. 1977{\natexlab{a}}, \actaa, 27, 203

\bibitem[{{Dziembowski}(1977{\natexlab{b}})}]{Dziembowski1977b}
{Dziembowski}, W. 1977{\natexlab{b}}, \actaa, 27, 95

\bibitem[{{Dziembowski} {et~al.}(2001){Dziembowski}, {Gough}, {Houdek}, \&
  {Sienkiewicz}}]{Dziembowski2001}
{Dziembowski}, W.~A., {Gough}, D.~O., {Houdek}, G., \& {Sienkiewicz}, R. 2001,
  \mnras, 328, 601

\bibitem[{{Edelmann} {et~al.}(2019){Edelmann}, {Ratnasingam}, {Pedersen},
  {Bowman}, {Prat}, \& {Rogers}}]{Edelmann2019}
{Edelmann}, P.~V.~F., {Ratnasingam}, R.~P., {Pedersen}, M.~G., {et~al.} 2019,
  \apj, 876, 4

\bibitem[{{Eggenberger} {et~al.}(2019{\natexlab{a}}){Eggenberger}, {Buldgen},
  \& {Salmon}}]{Eggenberger2019}
{Eggenberger}, P., {Buldgen}, G., \& {Salmon}, S.~J.~A.~J. 2019{\natexlab{a}},
  \aap, 626, L1

\bibitem[{{Eggenberger} {et~al.}(2019{\natexlab{b}}){Eggenberger}, {Deheuvels},
  {Miglio}, {Ekstr{\"o}m}, {Georgy}, {Meynet}, {Lagarde}, {Salmon}, {Buldgen},
  {Montalb{\'a}n}, {Spada}, \& {Ballot}}]{Eggenberger2019b}
{Eggenberger}, P., {Deheuvels}, S., {Miglio}, A., {et~al.} 2019{\natexlab{b}},
  \aap, 621, A66

\bibitem[{{Fossat} {et~al.}(2017){Fossat}, {Boumier}, {Corbard}, {Provost},
  {Salabert}, {Schmider}, {Gabriel}, {Grec}, {Renaud}, {Robillot},
  {Roca-Cort{\'e}s}, {Turck-Chi{\`e}ze}, {Ulrich}, \& {Lazrek}}]{Fossat2017}
{Fossat}, E., {Boumier}, P., {Corbard}, T., {et~al.} 2017, \aap, 604, A40

\bibitem[{{Fossat} \& {Schmider}(2018)}]{Fossat2018}
{Fossat}, E. \& {Schmider}, F.~X. 2018, \aap, 612, L1

\bibitem[{{Gabriel} {et~al.}(1995){Gabriel}, {Grec}, {Charra}, {Robillot},
  {Roca Cort{\'e}s}, {Turck-Chi{\`e}ze}, {Bocchia}, {Boumier}, {Cantin},
  {Cesp{\'e}des}, {Cougrand }, {Cr{\'e}tolle}, {Dam{\'e}}, {Decaudin},
  {Delache}, {Denis}, {Duc}, {Dzitko}, {Fossat}, {Fourmond}, {Garc{\'\i}a},
  {Gough}, {Grivel}, {Herreros}, {Lagard{\`e}re}, {Moalic}, {Pall{\'e}},
  {P{\'e}trou}, {Sanchez}, {Ulrich}, \& {van der Raay}}]{Gabriel1995}
{Gabriel}, A.~H., {Grec}, G., {Charra}, J., {et~al.} 1995, \solphys, 162, 61

\bibitem[{{Garc{\'{\i}}a} {et~al.}(2007){Garc{\'{\i}}a}, {Turck-Chi{\`e}ze},
  {Jim{\'e}nez-Reyes}, {Ballot}, {Pall{\'e}}, {Eff-Darwich}, {Mathur}, \&
  {Provost}}]{Garcia2007}
{Garc{\'{\i}}a}, R.~A., {Turck-Chi{\`e}ze}, S., {Jim{\'e}nez-Reyes}, S.~J.,
  {et~al.} 2007, Science, 316, 1591

\bibitem[{{Gehan} {et~al.}(2018){Gehan}, {Mosser}, {Michel}, {Samadi}, \&
  {Kallinger}}]{Gehan2018}
{Gehan}, C., {Mosser}, B., {Michel}, E., {Samadi}, R., \& {Kallinger}, T. 2018,
  \aap, 616, A24

\bibitem[{{Godart} {et~al.}(2009){Godart}, {Noels}, {Dupret}, \&
  {Lebreton}}]{Godart2009}
{Godart}, M., {Noels}, A., {Dupret}, M.-A., \& {Lebreton}, Y. 2009, \mnras,
  396, 1833

\bibitem[{{Gough}(1985)}]{Gough1985}
{Gough}, D.~O. 1985, in ESA Special Publication, Vol. 235, Future Missions in
  Solar, Heliospheric \& Space Plasma Physics, ed. E.~{Rolfe} \& B.~{Battrick},
  183

\bibitem[{{Kevorkian}(1961)}]{Kevorkian1961}
{Kevorkian}, J. 1961, PhD thesis, California Institute of Technology

\bibitem[{{Kiraga} {et~al.}(2005){Kiraga}, {Stepien}, \& {Jahn}}]{Kiraga2005}
{Kiraga}, M., {Stepien}, K., \& {Jahn}, K. 2005, \actaa, 55, 205

\bibitem[{{Kosovichev}(2011)}]{Kosovichev2011}
{Kosovichev}, A.~G. 2011, {Advances in Global and Local Helioseismology: An
  Introductory Review}, ed. J.-P. {Rozelot} \& C.~{Neiner}, Vol. 832, 3

\bibitem[{{Kumar} {et~al.}(1996){Kumar}, {Quataert}, \& {Bahcall}}]{Kumar1996}
{Kumar}, P., {Quataert}, E.~J., \& {Bahcall}, J.~N. 1996, \apjl, 458, L83

\bibitem[{{Lecoanet} \& {Quataert}(2013)}]{Lecoanet2013}
{Lecoanet}, D. \& {Quataert}, E. 2013, \mnras, 430, 2363

\bibitem[{{Ledoux}(1951)}]{Ledoux1951}
{Ledoux}, P. 1951, \apj, 114, 373

\bibitem[{{Lighthill}(1978)}]{Lighthill1978}
{Lighthill}, J. 1978, {Waves in fluids}

\bibitem[{{Lopes} \& {Turck-Chi{\`e}ze}(2014)}]{Lopes2014}
{Lopes}, I. \& {Turck-Chi{\`e}ze}, S. 2014, \apjl, 792, L35

\bibitem[{{Marques} {et~al.}(2013){Marques}, {Goupil}, {Lebreton}, {Talon},
  {Palacios}, {Belkacem}, {Ouazzani}, {Mosser}, {Moya}, {Morel}, {Pichon},
  {Mathis}, {Zahn}, {Turck-Chi{\`e}ze}, \& {Nghiem}}]{Marques2013}
{Marques}, J.~P., {Goupil}, M.~J., {Lebreton}, Y., {et~al.} 2013, \aap, 549,
  A74

\bibitem[{{Mussack} \& {D{\"a}ppen}(2011)}]{Mussack2011}
{Mussack}, K. \& {D{\"a}ppen}, W. 2011, \apj, 729, 96

\bibitem[{{Pieri} {et~al.}(2016){Pieri}, {Falasca}, {von Hardenberg}, \&
  {Provenzale}}]{Pieri2016}
{Pieri}, A.~B., {Falasca}, F., {von Hardenberg}, J., \& {Provenzale}, A. 2016,
  Physics Letters A, 380, 1363

\bibitem[{{Pin{\c c}on} {et~al.}(2016){Pin{\c c}on}, {Belkacem}, \&
  {Goupil}}]{Pincon2016}
{Pin{\c c}on}, C., {Belkacem}, K., \& {Goupil}, M.~J. 2016, \aap, 588, A122

\bibitem[{{Pin{\c c}on} {et~al.}(2017){Pin{\c c}on}, {Belkacem}, {Goupil}, \&
  {Marques}}]{Pincon2017}
{Pin{\c c}on}, C., {Belkacem}, K., {Goupil}, M.~J., \& {Marques}, J.~P. 2017,
  \aap, 605, A31

\bibitem[{Pin{\c c}on(2017)}]{mathese}
Pin{\c c}on, C. 2017, Theses, {PSL Research University}

\bibitem[{{Pratt} {et~al.}(2017){Pratt}, {Baraffe}, {Goffrey}, {Constantino},
  {Viallet}, {Popov}, {Walder}, \& {Folini}}]{Pratt2017}
{Pratt}, J., {Baraffe}, I., {Goffrey}, T., {et~al.} 2017, \aap, 604, A125

\bibitem[{{Rempel}(2004)}]{Rempel2004}
{Rempel}, M. 2004, \apj, 607, 1046

\bibitem[{{Rieutord} \& {Zahn}(1995)}]{Rieutord1995}
{Rieutord}, M. \& {Zahn}, J.-P. 1995, \aap, 296, 127

\bibitem[{{Rogers} {et~al.}(2006){Rogers}, {Glatzmaier}, \&
  {Jones}}]{Rogers2006b}
{Rogers}, T.~M., {Glatzmaier}, G.~A., \& {Jones}, C.~A. 2006, \apj, 653, 765

\bibitem[{{Rogers} {et~al.}(2013){Rogers}, {Lin}, {McElwaine}, \&
  {Lau}}]{Rogers2013}
{Rogers}, T.~M., {Lin}, D.~N.~C., {McElwaine}, J.~N., \& {Lau}, H.~H.~B. 2013,
  \apj, 772, 21

\bibitem[{{Scherrer} \& {Gough}(2019)}]{Scherrer2019}
{Scherrer}, P.~H. \& {Gough}, D.~O. 2019, \apj, 877, 42

\bibitem[{{Schunker} {et~al.}(2018){Schunker}, {Schou}, {Gaulme}, \&
  {Gizon}}]{Schunker2018}
{Schunker}, H., {Schou}, J., {Gaulme}, P., \& {Gizon}, L. 2018, \solphys, 293,
  95

\bibitem[{{Severnyi} {et~al.}(1976){Severnyi}, {Kotov}, \&
  {Tsap}}]{Severnyi1976}
{Severnyi}, A.~B., {Kotov}, V.~A., \& {Tsap}, T.~T. 1976, \nat, 259, 87

\bibitem[{{Shibahashi}(1979)}]{Shibahashi1979}
{Shibahashi}, H. 1979, \pasj, 31, 87

\bibitem[{{Stein} \& {Nordlund}(1998)}]{Stein1998}
{Stein}, R.~F. \& {Nordlund}, A. 1998, \apj, 499, 914

\bibitem[{{Tassoul}(1980)}]{Tassoul1980}
{Tassoul}, M. 1980, \apj, 43, 469

\bibitem[{{Thompson} {et~al.}(2003){Thompson}, {Christensen-Dalsgaard},
  {Miesch}, \& {Toomre}}]{Thompson2003}
{Thompson}, M.~J., {Christensen-Dalsgaard}, J., {Miesch}, M.~S., \& {Toomre},
  J. 2003, \araa, 41, 599

\bibitem[{{Thomson} {et~al.}(1995){Thomson}, {Maclennan}, \&
  {Lanzerotti}}]{Thomson1995}
{Thomson}, D.~J., {Maclennan}, C.~G., \& {Lanzerotti}, L.~J. 1995, \nat, 376,
  139

\bibitem[{{Townsend}(1966)}]{Townsend1966}
{Townsend}, A.~A. 1966, Journal of Fluid Mechanics, 24, 307

\bibitem[{{Turck-Chi{\`e}ze} {et~al.}(2004){Turck-Chi{\`e}ze}, {Garc{\'{\i}}a},
  {Couvidat}, {Ulrich}, {Bertello}, {Varadi}, {Kosovichev}, {Gabriel},
  {Berthomieu}, {Brun}, {Lopes}, {Pall{\'e}}, {Provost}, {Robillot}, \& {Roca
  Cort{\'e}s}}]{TC2004}
{Turck-Chi{\`e}ze}, S., {Garc{\'{\i}}a}, R.~A., {Couvidat}, S., {et~al.} 2004,
  \apj, 604, 455

\bibitem[{{Turner}(1986)}]{Turner1986}
{Turner}, J.~S. 1986, Journal of Fluid Mechanics, 173, 431

\bibitem[{{Ulrich} {et~al.}(2000){Ulrich}, {Boumier}, {Robillot},
  {Garc{\'\i}a}, {Roca Cort{\'e}s}, \& {Henney}}]{Ulrich2000}
{Ulrich}, R.~K., {Boumier}, P., {Robillot}, J.~M., {et~al.} 2000, \aap, 364,
  816

\bibitem[{{Unno} {et~al.}(1989){Unno}, {Osaki}, {Ando}, {Saio}, \&
  {Shibahashi}}]{Unno1989}
{Unno}, W., {Osaki}, Y., {Ando}, H., {Saio}, H., \& {Shibahashi}, H. 1989,
  {Nonradial oscillations of stars}

\bibitem[{{von Hardenberg} {et~al.}(2008){von Hardenberg}, {Parodi}, {Passoni},
  {Provenzale}, \& {Spiegel}}]{Hardenberg2008}
{von Hardenberg}, J., {Parodi}, A., {Passoni}, G., {Provenzale}, A., \&
  {Spiegel}, E.~A. 2008, Physics Letters A, 372, 2223

\bibitem[{{Zahn}(1991)}]{Zahn1991}
{Zahn}, J.-P. 1991, \aap, 252, 179

\end{thebibliography}


\appendix
\renewcommand\appendixname{}

\section{Energy spectrum of plume-induced gravity modes}
\label{derivation}

In the following, we aim to derive the energy spectrum of asymptotic gravity modes generated by penetrative convection. The mathematical derivation largely uses and sometimes extends the analysis of stellar oscillations presented in \cite{Unno1989}, as well as the model of \cite{Pincon2016}.

\subsection{Forced oscillation equations}
\label{forced momentum}

Following the model of \cite{Pincon2016}, the linearized momentum, continuity, Poisson's equations and the equation of state\footnote{The Lagrangian variation of the mean molecular weight is usually neglected because the microscopic diffusion and nuclear timescales are supposed to be much longer than the oscillation timescale.} read  
\begin{align}
&\rho~ \partial _t^2 \vec{\xi} +\vec{\nabla} p^\prime- \rho^\prime \vec{g}+\rho \vec{\nabla}\psi^\prime = -\vec{\nabla} \cdot (\rho \vec{\mathcal{V}}_{\rm p}\otimes \vec{\mathcal{V}}_{\rm p})\label{momentum eq} \\
&\delta \rho+\rho \vec{\nabla}\cdot \vec{\xi}=0 \label{continuity eq}\\
&\nabla^2 \psi^\prime = 4\pi G \rho^\prime \label{Poisson eq} \\
&\frac{\delta \rho}{\rho}=\frac{1}{\Gamma_1}\frac{\delta p}{p}-\varv_T \frac{\delta S}{c_p} \label{EOS eq}\; ,
\end{align}
where $p$ and $\vec{g}$ are the equilibrium pressure and gravitational acceleration, respectively, $\psi^\prime$ is the perturbation of the gravitational potential, $\Gamma_1$ is the first adiabatic index, $G$ is the gravitation constant, $\varv_T=-(\partial \ln \rho/\partial \ln T)_p$, and all the other quantities are introduced in \sectionname{}~\ref{forced equation}. In the latter equations, $X^\prime$ and $\delta X$ denotes the Eulerian and Lagrangian perturbations of the quantity $X$, respectively.
Using the usual solution of the Poisson's equation
\algn{
\psi^\prime (\vec{r},t) = - \int_{V} G \frac{\rho^\prime (\vec{x},t) }{|\vec{r}-\vec{x}|} \dd V \; ,
\label{Poisson sol}
}
where $V$ is the stellar volume beyond which the stellar density vanishes,
as well as \eqs{continuity eq}{EOS eq}, it is possible to rewrite \eq{momentum eq} in the form of \eq{complete momentum eq}
in which
\algn{
\vec{\mathcal{L}}^{\rm nad}\left(H\frac{\delta S}{c_p} \right)= \frac{1}{\rho}\vec{\nabla}\left(\Gamma_1\varv_T  p \frac{\delta S}{c_p} \right) \; ,
\label{L^nad}
}
with $H=-(\dd r / \dd \ln T)$ the temperature scale height and $T$ the equilibrium temperature, and where \smash{$\vec{\mathcal{L}}^{\rm ad}$} is the adiabatic differential linear operator defined as \citep[e.g.,][see Chap. 14.3 for details]{Chandra1964,Unno1989}
\algn{
\vec{\mathcal{L}}^{\rm ad}\left( \vec{\xi}\right)&=\frac{\vec{\nabla} p}{\rho^2} \vec{\nabla} \cdot\left(\rho \vec{\xi}\right) 
-\frac{1}{\rho} \vec{\nabla} \left ( \vec{\xi}\cdot \vec{\nabla} p\right)
-\frac{1}{\rho} \vec{\nabla} \left ( \rho c^2\vec{\nabla}\cdot\vec{\xi}  \right)\nonumber \\
&+ \vec{\nabla} \left( \int_{V} G \frac{\vec{\nabla}_{\vec{x}}\cdot\left[\rho(\vec{x}) \vec{\xi}(\vec{x}) \right] }{|\vec{r}-\vec{x}|} \dd \vec{x}^3 \right) \; ,
\label{L^ad}
}
with $c^2=\Gamma_1 p/\rho$ the squared sound speed. We note that $H$ has been introduced in \eq{L^nad} for $\smash{\vec{\mathcal{L}}^{\rm nad}}$ to represent the same physical quantity as $\smash{\vec{\mathcal{L}}^{\rm ad}}$ (i.e., a squared frequency).
Equation~(\ref{complete momentum eq}) has to be completed by the energy equation that specifies the evolution of $\delta S$. Neglecting the contribution of the convective flux for asymptotic gravity modes \citep{Belkacem2009} and the nuclear energy production rate in the considered layers,  it reads
\algn{
T\partial_t \delta S = -~ \delta \left(\frac{1}{\rho} \vec{\nabla}\cdot \vec{\mathcal{F}}_{\rm R} \right) \; ,
\label{energy eq}
}
where $\vec{\mathcal{F}}_{\rm R}$ is the radiative flux.

\subsection{Mode amplitude}

\subsubsection{Energy equation in the diffusion approximation}
\label{diffusion}

At this stage, the oscillation equations are represented by \eqs{complete momentum eq}{energy eq}. To go further, we thus need to express the Lagrangian perturbation of the radiative flux. In the diffusion approximation, the radiative flux is equal to
\algn{
\vec{\mathcal{F}}_{\rm R}= -\rho c_p K_{\rm rad} \vec{\nabla}T=-\frac{16\sigma_{\rm SB} T^3}{3\rho \kappa} \vec{\nabla}T \; ,
}
where $K_{\rm rad}$ is the radiative diffusivity, $\sigma_{\rm SB} $ is the Stefan-Boltzmann's constant and $\kappa$ is the Rosseland's mean opacity. Perturbing this latter equation and replacing $\delta \rho/\rho$ by $-\vec{\nabla}\cdot \vec{\xi}$, \eq{energy eq} can be rewritten
\algn{
\partial_t \left( \frac{\delta S}{c_p} \right) =&-\frac{1}{\rho c_p T} \vec{\nabla}\cdot \left[-F_{\rm R}H\vec{\nabla}\left(\frac{\delta T}{T} \right) +\left(4\frac{\delta T}{T}+\vec{\nabla}\cdot \vec{\xi}-\frac{\delta \kappa}{\kappa} \right)\vec{F}_{\rm R}\right]\nonumber\\
&+\frac{1}{\rho c_p T} \vec{\nabla}\cdot \left[\vec{\nabla} \left(\vec{F}_{\rm R}\cdot \vec{\xi}\right)-\vec{\xi}\left(\vec{\nabla}\cdot \vec{F}_{\rm R} \right)\right] \; ,
\label{energy perturb eq}
}
where $\vec{F}_{\rm R}=F_{\rm R} \vec{e}_r$ is the radial equilibrium radiative flux. We note that the right-hand side of \eq{energy perturb eq} is equivalent to the right-hand side of  Eq.~(21.14) of \cite{Unno1989} with \smash{$\delta \varepsilon_N=0$.} Considering the equation of state $T(S,\rho)$ and the opacity table $\kappa(T,\rho)$ while using the continuity equation, it becomes
\algn{
\frac{\delta T}{T}&=\gamma \frac{\delta S}{c_p} -\Gamma_1 \nabla_{\rm ad} \vec{\nabla}\cdot \vec{\xi} \label{T EOS}\\
\frac{\delta \kappa}{\kappa}&=\gamma \kappa_T  \frac{\delta S}{c_p}-\left(\kappa_T\Gamma_1 \nabla_{\rm ad}+\kappa_\rho\right)\vec{\nabla}\cdot \vec{\xi} \label{kappa EOS}\; ,
}
where $\gamma=c_p/c_V$, with $c_p$ and $c_V$ the specific heat capacities at constant pressure and volume, respectively, $\nabla_{\rm ad}$ is the adiabatic temperature gradient, $\kappa_T=(\partial \ln\kappa/\partial \ln T)_\rho$ and $\kappa_\rho=(\partial \ln\kappa/\partial \ln \rho)_T$. Using \eqs{T EOS}{kappa EOS}~in \eq{energy perturb eq}, the evolution of the Lagrangian perturbation of entropy is ruled by
\algn{
\partial_t \left( \frac{\delta S}{c_p} \right)=\frac{1}{t_{\rm R}} \left[ \mathcal{L}^{{\rm nad}1}\left( \frac{\vec{\xi}}{H}\right)+\mathcal{L}^{{\rm nad}2}\left( \frac{\delta S}{c_p}\right)\right] \; ,
\label{energy perturb eq 2 appendix}
}
where we have introduced the local radiative thermal timescale
\algn{
t_{\rm R} = \frac{\rho c_p T H}{F_{\rm R}}\; ,
\label{t_th}
}
and where \smash{$\mathcal{L}^{{\rm nad}1}$ and $\mathcal{L}^{{\rm nad}2}$} are two linear operators that involves derivatives with respect to the normalized variable $r/H$ and whose expressions can be readily deduced from \eqss{energy perturb eq}{kappa EOS}.

\subsubsection{Local scaling of the oscillation equations}
\label{scaling}

Before going further, it is instructive to express locally the oscillation equations in a dimensionless form. In the following, we focus on a mode with a characteristic angular frequency and a local wavelength denoted with $\sigma$ and $\lambda(r,\sigma)$, respectively. The dynamical timescale is thus defined as \smash{$t_{\rm dyn}\equiv\sigma^{-1}$.} 
First, according to \eq{eigenvalue}, the local norm of the \smash{$\vec{\mathcal{L}}^{\rm ad}$} operator\footnote{In this paper, the local norm of a linear operator $\mathcal{L}$ acting on a vector $X(\vec{r})$ in the vicinity of a point $\vec{r}_0$ is defined as
\algn{ \mathcal{N}_X(\vec{r}_0)=\max_{\vec{r}\in C_0} \left( \frac{\left|\mathcal{L}[X(\vec{r})]\right|}{ \left|X(\vec{r})\right| }\right)\; ,} where $(|\cdot|)$ is the modulus and $C_0$ represents the volume of the sphere of center $\vec{r}_0$ with a radius equal to the local characteristic wavelength $\lambda(r_0)$ of the vector $X$, excluding the nodes where $X(\vec{r})= 0$.} 
is equal to $\sigma^{2}$ for such a mode. Owing to the place of \smash{$\vec{\mathcal{L}}^{\rm nad}$} in \eq{complete momentum eq}, it seems reasonable to assume that its norm scales also as $\sigma^{2}$. This hypothesis is checked a posteriori in \sectionname{}~\ref{damping rate}. Equation~(\ref{complete momentum eq}) can thus be rewritten as
\algn{
\partial _\tau^2 \vec{\xi} +\vec{\tilde{\mathcal{L}}}^{\rm ad}\left( \vec{\xi}\right) +\vec{\tilde{\mathcal{L}}}^{\rm nad}\left(H \frac{\delta S}{c_p} \right)=-\frac{1}{\rho \sigma^2}\vec{\nabla} \cdot (\rho \vec{\mathcal{V}}_{\rm p}\otimes \vec{\mathcal{V}}_{\rm p}) \; ,
\label{scaled momentum eq}
}
where we have defined $\tau=t/t_{\rm dyn}$ and the differential operators \smash{$\vec{\tilde{\mathcal{L}}}^{\rm ad} =\sigma^{-2}\vec{\mathcal{L}}^{\rm ad}$} and \smash{$\vec{\tilde{\mathcal{L}}}^{\rm nad} =\sigma^{-2}\vec{\mathcal{L}}^{\rm nad}$} such as their norms remain on the order of unity.
Second, regarding the energy equation, we firstly note that, owing to the incompressible character of the asymptotic low-frequency gravity modes \cite[e.g.,][]{Dintrans2001},
\algn{\vec{\nabla}\cdot \vec{\xi} = \mathcal{O}\Big(|\vec{\xi}|/H \Big) \; ,
\label{incomp}
}
where the big-$\mathcal{O}$ Bachmann-Landau's notation is introduced. As $\mathcal{L}^{{\rm nad}1}$ and $\mathcal{L}^{{\rm nad}2}$ in \eq{energy perturb eq 2 appendix} involve third-order  and second-order derivatives at most with respect to $r/H$ of \smash{$\vec{\nabla}\cdot \vec{\xi}$} and \smash{$\delta S/c_p$}, respectively, their norms thus scale as $(H^2/\lambda^2$) at most since $\lambda\lesssim H$ and \eq{incomp} are met within the asymptotic limit. Equation~(\ref{energy perturb eq 2 appendix}) can hence be locally expressed as
\algn{
\partial_\tau \left( \frac{\delta S}{c_p} \right)=\varepsilon_{\sigma} \left[ \tilde{\mathcal{L}}^{{\rm nad}1}\left( \frac{\vec{\xi}}{H}\right)+\tilde{\mathcal{L}}^{{\rm nad}2}\left( \frac{\delta S}{c_p}\right)\right]  \; ,
\label{scaled energy eq}
}
In this latter equation, we have defined
\algn{\varepsilon_{\sigma}(r)\equiv\frac{t_{\rm dyn}}{t_{\rm damp}(r,\sigma)}~~~~\mbox{and}~~~~\tilde{\mathcal{L}}^{{\rm nad}k}\equiv\frac{\lambda^2(r,\sigma)}{H^2(r)}\mathcal{L}^{{\rm nad}k}\; ,}
in such a way that the norms of $\tilde{\mathcal{L}}^{{\rm nad}k}$ for $k=1,2$ remain on the order of unity too, and where we can identify the local damping timescale as
\algn{t_{\rm damp}(r,\sigma)\equiv\frac{\lambda^2(r,\sigma)}{H^2(r)}t_{\rm R}(r) \label{t_damp}\; .}
 %

\subsubsection{Globally quasi-adiabatic oscillations}
\label{quasi-adiabatic}

At this point, we focus on a component $(n,\ell,m)$ with the amplitude $a_{n\ell m}(t)$ and the frequency $\omega_{n\ell m}$, and fix $\sigma=\omega_{n\ell m}$ and $\varepsilon_\sigma\equiv\varepsilon_{n\ell m}$. According to \eq{scaled energy eq}, $\delta S$ locally scales as
\algn{
\frac{\delta S}{c_p} = \mathcal{O}\left( \varepsilon_{n\ell m}\frac{|\vec{\xi}_{n\ell m}|}{H}\right) \; .
\label{delta S scaling}
}
In a given frequency range around $\omega_{n\ell m}$, the damping timescale is expected to be much longer than the oscillation period almost everywhere in the star.
As a result, $\varepsilon_{n\ell m}\ll1$ and \smash{$\delta S/c_p \ll |\vec{\xi}_{n\ell m}|/H$.}
In the limit of $\varepsilon_{n\ell m}\rightarrow 0$, $\delta S\rightarrow 0$ and the problem locally tends toward the adiabatic case. Nevertheless, there obviously exists a very thin near-surface layer where $\varepsilon_{n\ell m} \gg1$ and the quasi-adiabatic approximation locally fails, so that the ordering in \eq{delta S scaling} is not met. Therefore, in order to reason from a global point of view, we define the global damping timescale as the inverse of the harmonic mean of the local damping timescale throughout the star weighted by the local mode energy, that is,
\algn{
T_{\rm damp}^{-1}   \equiv \frac{1}{\mathcal{M}_{n\ell m}} \int_V t_{\rm damp}^{-1} \; \rho \vec{\xi}_{n\ell m} \cdot \vec{\xi}_{n\ell m}^\star  \dd V \; ,
\label{T_damp}
}
where  $T_{\rm damp}$ has to be related to the local wavelength of $\vec{\xi}_{n\ell m}$ through \eqs{t_damp}{T_damp}, and where $\mathcal{M}_{n\ell m}$ is the mode mass defined in \eq{orthogonality}.
In the following, we thus assume that the oscillations are globally quasi-adiabatic, that is,
\algn{
\delta_{n\ell m} \equiv t_{\rm dyn}/T_{\rm damp} \ll 1 \; .
\label{global adiabatic}
}
For convenience, we also define the quantity
\algn{
\chi_{n\ell m}= \mathcal{M}_{n\ell m} \delta_{n\ell m} \; .
}
The global quasi-adiabatic approximation expressed by \eq{global adiabatic} will greatly ease the derivation of the mode amplitude, as we will see.

Taking the derivative of \eq{scaled momentum eq} with respect to $\tau$, injecting \eqs{field expansion}{scaled energy eq}~into the obtained expression, and computing the inner product with $\vec{\xi}_{n\ell m}^\star$, we obtain a differential equation ruling the mode amplitude $a_{n\ell m}(\tau)$. We point out that, owing to the spherical symmetry of the star, the inner product with $\vec{\xi}_{n\ell m}^\star$ selects the angular degree $\ell$ and the azimuthal number $m$ in the expansion of the perturbations onto the orthonormal basis of the spherical harmonics (e.g., see the definitions of the differential operators to be convinced). Therefore, the subscript $\ell m$ is often dropped to simplify the notation in what follows. The amplitude equation finally reads, at first order in $\delta_{n}$,
\algn{
\partial_\tau^3 a_{n} +\partial_\tau a_{n}-2 \delta_{n}\left( \tilde{\eta}_{nn}a_{n}+ \sum_{n^\prime\ne n}\tilde{\eta}_{n n^\prime}a_{n^\prime}\right)= \partial_\tau \widetilde{\mathcal{F}}_{n}+\mathcal{O}(\delta_n^2) \; ,
\label{amplitude eq}
}
where
\algn{
\tilde{\eta}_{n n^\prime}&=-\frac{1}{2 \chi_{n\ell m}}\int_{V}\rho \vec{\xi}_{n\ell m}^\star \cdot \tilde{\vec{\mathcal{L}}}^{\rm nad}\left[H\varepsilon_{n^\prime\ell m}\tilde{\vec{\mathcal{L}}}^{\rm nad1}\left(\frac{\vec{\xi}_{n^\prime\ell m}}{H}\right)\right] \dd V\label{mode coupling}\\
\widetilde{\mathcal{F}}_{n} &=-\frac{1}{\omega_{n\ell m}^2\mathcal{M}_{n\ell m}}\int_{V} \vec{\nabla} \cdot (\rho \vec{\mathcal{V}}_{\rm p}\otimes \vec{\mathcal{V}}_{\rm p})\cdot \vec{\xi}_{n\ell m}^\star \dd V \label{mode driving} \; .
}
The second-order term in \eq{amplitude eq} is related to the Lagrangian perturbation of entropy and results from the scaling in \eq{delta S scaling} and the global quasi-adiabatic hypothesis in \eq{global adiabatic}. Besides, the product $\delta_{n}\tilde{\eta}_{nn}$ stands for the damping rate (in terms of the $\tau$ variable). In contrast, the sum in the brackets of the left-hand side of \eq{amplitude eq} encapsulates coupling terms between the components $n$ and $n^\prime\ne n$. Such coupling terms result, by construction, from the expansion onto the eigenfunctions of the \smash{$\vec{\mathcal{L}}^{\rm ad}$} operator, as they are not natural solutions of the non-adiabatic problem. When writing \eq{amplitude eq}, we have implicitly assumed $\tilde{\eta}_{n n^\prime}= \mathcal{O}(1)$. Indeed, as shown in \eq{mode coupling}, the integral in the numerator is at most of the order of $\chi_{n\ell m}$ since the norms of the operators are on the order of unity. Moreover, this integral correspond to the inner product between an oscillating radial function, with $n$ radial nodes and a characteristic wavelength $\lambda_n$, and another oscillating radial function, with about $n^\prime$ radial nodes and a characteristic wavelength $\lambda_{n^\prime}$. For $n^\prime \sim n$, we have therefore $|\tilde{\eta}_{n n^\prime}|\sim 1$, whereas for $n^\prime \gg n$ or $n^\prime \ll n$, both oscillating functions are incoherent with each other and the inner product vanishes, that is, $|\tilde{\eta}_{n n^\prime}|\ll 1$. This justifies $\tilde{\eta}_{nn^\prime} = \mathcal{O}(1)$.

\subsubsection{Two-timing analysis}
\label{two-timing}

As shown by \eq{amplitude eq}, the evolution of the mode amplitude is ruled by a fast dynamical timescale, $T_0=\tau=t/t_{\rm dyn}$, and a slow damping timescale,  $T_1=\delta_n\tau=t/T_{\rm damp}$. Using a two-timing perturbation method therefore seems judicious to solve analytically this equation \citep[e.g.,][]{Kevorkian1961}. Such a method can provide us with a uniformly valid solution up to timescale of order of $\mathcal{O}(1/\delta_n)$ in terms of the independent variable $\tau$, or equivalently $\mathcal{O}(T_{\rm damp})$ in terms of the dependent variable $t$. 

\paragraph{Homogeneous equation.} Taking advantage of the linearity of the problem, we search in a first step for a general solution of the homogeneous amplitude equation. For all $n$, we start from the two-timing perturbation ansatz, that is,
\algn{
a_n(\tau)=a_n^0(T_0,T_1)+\delta_n a_n^1(T_0,T_1)+\mathcal{O}(\delta_n^2) \; ,
\label{ansatz}
}
keeping in mind that $\tau$, $T_0$ and $T_1$ also depends on $n$ through the scaling by $1/\omega_{n\ell m}$.
Injecting \eq{ansatz} into \eq{amplitude eq} and using $\partial_\tau=\partial_{T_0}+\delta_n\partial_{T_1}$ via the chain rule, the homogeneous equation reads at leading order in $\delta_n$
\algn{
\partial_{T_0}^3 a_n^0 +\partial_{T_0} a_n^0=0 \label{zero order}\; ,
}
 and at first order in $\delta_n$,
 \algn{
 \partial_{T_0}^3 a_n^1  +\partial_{T_0} a_n^1 + \underbrace{\left( 3  \partial_{T_0}^2\partial_{T_1} a_n^0+  \partial_{T_1} a_n^0 -2\tilde{\eta}_{nn} a_n^0\right)}_{\mbox{Secular terms}}=
 2\sum_{\substack{n^\prime\ne n}}\tilde{\eta}_{n n^\prime}a_{n^\prime}^0 \label{first order}\; .
 }
The general solution of \eq{zero order} then reads
 \algn{
 a_n^0(T_1,T_0)=\alpha_n(T_1) e^{iT_0} +\beta_n(T_1) e^{-iT_0} \; ,
 \label{zero solution}
 }
where $\alpha_n(T_1)$ and $\beta_n(T_1)$ are two functions to be determined with the first-order equation.
Using \eq{zero solution} in \eq{first order}, it is straightforward to show (e.g., via the method of variation of parameters) that, if non-null, the terms in brackets will induce terms proportional to \smash{$\tau e^{\pm i\tau}$} in the expression of $a_n^1$, up to a factor of the order of $\alpha_n$ and $\beta_n$. This fact will make $a_n^1$ of the order of $\mathcal{O}(a_n^0/\delta_n)$ up to a timescale of the order of $\mathcal{O}(1/\delta_n)$ and, therefore, would break the perturbation ansatz in \eq{ansatz}. Therefore, to obtain a uniformly valid expression up to such timescales, the terms in brackets have to vanish. This condition leads to
\algn{
\alpha_n(T_1)=A_ne^{-\tilde{\eta}_{nn}T_1}~~~~\mbox{and}~~~~\beta_n(T_1)=B_ne^{-\tilde{\eta}_{nn}T_1} \; ,
\label{alpha beta}
}
with $A_n$ and $B_n$ two arbitrary constants.
As a consequence, the amplitude $a_n^1$ is ruled only by the coupling term in the right-hand side of \eq{first order}, which depends on the leading-order amplitudes $a_{n^\prime}^0$ with $n^\prime \neq n$. For $n^\prime \neq n$, it is obvious that 
the amplitudes follow the same zeroth-order and first-order equations as in \eqs{zero order}{first order}, except that, because of the considered scaling, $T_k$ for $k=0$ or $1$ in these equations are replaced by \smash{$T_k^\prime=\omega_{n^\prime} t=\varpi T_k$} with $\varpi=(\omega_{n^\prime}/\omega_n)$. As a result, we get
\algn{
a_{n^\prime}^0(T_1,T_0)=A_{n^\prime}e^{-\tilde{\eta}_{n^\prime n^\prime}\varpi T_1} e^{i\varpi T_0} +B_{n^\prime}e^{-\tilde{\eta}_{n^\prime n^\prime }\varpi T_1} e^{-i\varpi T_0} \; ,
\label{a_n^prime}
}
where $A_{n^\prime}$ and $B_{n^\prime}$ are two arbitrary constants. According to \eqss{first order}{alpha beta}, we thus find that $a_{n}^1$ is ruled at leading order by
\algn{
 \partial_{T_0}^3 a_n^1  +\partial_{T_0} a_n^1 =
 2\sum_{\substack{n^\prime\ne n}}\tilde{\eta}_{n n^\prime}a_{n^\prime}^0 \label{first order a_n^1}\; .
 }
Using again the method of variation of parameters, we can show that the amplitude $a_n^1$ is equal to a series of terms as a function of $n^\prime$ that are $\mathcal{O}(\tilde{\eta}_{n n^\prime} a_{n^\prime}^0/|\varpi-1|$). At this point, we assume that the magnitude of \smash{$a_{n^\prime}^0$} slowly varies with $n^\prime$, so that it is \smash{$\mathcal{O}(a_n^0)$} for \smash{$n^\prime \sim n$}; this will be checked a posteriori (see the resulting velocity amplitudes in \figurename{}~\ref{v_app_1}). Provided $\tilde{\eta}_{n n^\prime}=\mathcal{O}(1)$ for \smash{$n\sim n^\prime$} and $\tilde{\eta}_{n n^\prime}\ll 1$ otherwise, as justified in \sectionname{}~\ref{quasi-adiabatic}, the main contribution to $a_n^1$ results from the terms of the series associated with $n^\prime \sim n$; as a result, $a_n^1=\mathcal{O}(a_{n}^0\omega_n / \Delta \omega_n)$ with $\Delta \omega_n=|\omega_n -\omega_{n-1}|$. Within the asymptotic limit, $\omega_n / \Delta \omega_{n} \sim n \gg 1$ and thus $a_n^1 \gg a_n^0$ in order of magnitude. This highlights afterwards that, instead of $\delta_n$, it is more appropriate to scale the non-adiabatic perturbation of the mode amplitude by
\algn{\tilde{\delta}_{n}\equiv (\omega_n/\Delta \omega_n)~\delta_n\; ,}
in such a way that the perturbation ansatz becomes
\algn{
a_n(\tau)=a_n^0(T_0,T_1)+\tilde{\delta}_n \tilde{a}_n^1(T_0,T_1)+\mathcal{O}(\tilde{\delta}_n^2) \; ,
\label{ansatz 2}
}
and the new first-order perturbation $\tilde{a}_n^1 = \mathcal{O}(a_n^0)$. As a result, the relative error when considering $a_n \approx a_n^0$ up to a timescale of order $T_{\rm damp}$ is of the order of $\tilde{\delta}_n$.
We check a posteriori at the end of \sectionname{}~\ref{input} that $\tilde{\delta}_{n}\ll 1$ in the considered asymptotic frequency range. Therefore, all the previous perturbation developments, fortunately, hold true and the coupling terms do not stand for secular terms. These latter have thus no effect on the two-timing solution provided by $a_n^0$, and are just responsible for the relative error of the order of $\tilde{\delta}_n$.

In summary, provided that the mode amplitude slowly varies with the radial order, the solution of the homogeneous equation as a function of the variable $t$, which is uniformly valid up to a timescale of the order of $\mathcal{O}(T_{\rm damp})$, takes the form, according to \eqs{zero solution}{alpha beta}, of the adiabatic solution modulated by an exponential damping term, that is,
\algn{
a_{n}(t) \approx A_ne^{+i\omega_n t-\eta_n t}+ B_ne^{-i\omega_n t-\eta_n t} \; ,
\label{a_n homo}
}
where the relative error is of the order of $\tilde{\delta}_n$ and the damping rate is equal to
\algn{
\eta_n=\omega_n\delta_n \tilde{\eta}_{nn} \: ,
\label{eta_n}
}
with $\delta_n$ and $ \tilde{\eta}_{nn}$ defined in \sectionname{}~\ref{quasi-adiabatic}.

Physically speaking, we note that the parameter $\tilde{\delta}_n$ measures the coupling induced by the non-adiabatic effects between adjacent harmonics in the decomposition onto the eigenfunctions of the \smash{$\vec{\mathcal{L}}^{\rm ad}$} operator. To show this, it is possible to express this parameter in more sensible ways. On the one hand, the eigenfrequencies of the adiabatic asymptotic gravity modes follow at leading order \citep[e.g.,][]{Shibahashi1979,Tassoul1980}
\algn{
\omega_{n} \approx \frac{2 \pi}{n\Delta \Pi_\ell}\; , \;\mbox{with}\; \Delta \Pi_\ell = \frac{2\pi^2}{\sqrt{\ell(\ell+1)}} \left( \int_0^{r_{\rm b}} N \frac{\dd r}{r}\right)^{-1} \; ,
\label{quantization}
}
where $\Delta \Pi_\ell$ is the so-called period-spacing, $r_{\rm b}$ is the radius of the base of the convective zone, and $N$ is the Brunt-Väisälä frequency defined as
\algn{N^2 = g\left(\frac{1}{\Gamma_1}\deriv{\ln p}{r} - \deriv{\ln \rho}{r}   \right)\; .
\label{Brunt}}
Therefore, for $n\gg1$ in the asymptotic regime, we have
\algn{
\Delta \omega_n &\approx  \pi \left( \int_0^{r_{\rm b}} \frac{k_r}{\omega_{n}}\dd r \right)^{-1} \; \mbox{with}\; k_r\approx\frac{\sqrt{\ell(\ell+1)}}{r}\left(\frac{N}{\omega_{n}}-1\right)^{1/2} \; ,
\label{k_r}
}
where $k_r$ is the local radial wavenumber. As the radial group velocity of gravity waves is about equal to $ \omega_{n}/k_r$ \citep[e.g.,][]{Unno1989}, \eq{k_r} shows that \smash{$\Delta \omega_n \sim 1/\Delta t_{\rm core}$}, where $\Delta t_{\rm core}$ is the time spent by a wave energy ray of frequency $\omega_n$ to cross the radiative core. We thus obtain \smash{$\tilde{\delta}_n \sim \Delta t_{\rm core} / T_{\rm damp}$}. On the other hand, using \smash{$\eta_n \sim 1/T_{\rm damp}$} according to our scaling, we have \smash{$\tilde{\delta}_n \sim \eta_n /\Delta \omega_n$}. The hypothesis \eq{no coupling} is thus equivalent to assume that the width of each oscillation peak around an adiabatic eigenfrequency is much smaller than the frequency spacing between two consecutive radial orders $n$ and $n+1$ in the power oscillation spectrum.

\paragraph{Forced amplitude} In a second step, we search for the solution of the forced amplitude equation that is uniformly valid up to timescales of the order of $T_{\rm damp}$ and satisfies the initial value condition $a_{n\ell m}(-\infty) = 0$. Using the method of variation of parameters, it is straightforward to show that such a solution is
\algn{
a_{n}(t) \approx A_n(t)e^{+i\omega_n t-\eta_n t}+ B_n(t)e^{-i\omega_n t-\eta_n t}\; ,
\label{a_n forced 2}
}
where $A_n(t)$ is a function of time provided by \eq{a_n forced} and $B_n(t)= A_n(t)^\star$.
As a check, \eq{a_n forced 2} leads to
\algn{
\partial_\tau^3 a_n &=  \partial_\tau \widetilde{\mathcal{F}}_{n}-2 \delta_n \tilde{\eta}_{nn}a_{n}\widetilde{\mathcal{F}}_{n}+\mathcal{O}(\delta_n^2) \label{check 1}\\
\partial_\tau a_n  &= -2 \delta_n \tilde{\eta}_{nn}a_{n} \label{check 2} \; .
}
Injecting \eqs{check 1}{check 2}~ in \eq{amplitude eq}, we see that the secular term represented by $2 \eta_{nn} a_n$ cancel out and that the relative residual is of the order of $\tilde{\delta}_n$ at most, as shown in the previous paragraph.

\subsection{Mean mode energy within the asymptotic limit}

Within the asymptotic limit, it is possible to use in a good approximation the leading-order WKB analytical expressions of the eigenfunctions in order to express the mean mode energy in a simple but sensible way.

\subsubsection{General expression for the mean mode energy}
\label{general expression}

Owing to the equipartition of the specific kinetic and potential energies in the case of gravity waves \citep[e.g.,][]{Lighthill1978}, we define the mean total oscillation energy as
\begin{equation}
\left \langle E \right \rangle
=\lim_{T\rightarrow +\infty}\frac{1}{T} \int_{-T/2}^{+T/2} \left(\int_{V}\rho(r) \left|\partial_t \vec{\xi}(\vec{r},t)\right|^2 \dd V\right)\dd t \; ,
\label{mean wave energy}
\end{equation}
where ($\langle  \cdot \rangle$) is the (ensemble or time) mean operator and $V$ represents the stellar volume.

To express \eq{mean wave energy} in a simple way, the same reasoning as done in Sects.~2.2.1~and~2.2.2 of \cite{Pincon2016} can be used. While \cite{Pincon2016} considered progressive wave packets propagating toward the center of the star and never returning back upward, it is in contrast necessary in the present case to consider the modal structure of the oscillations. The only difference is that, instead of considering that the lifetime of the wave packet at a certain point of the radiative zone is of the order of $\tau_p$ as in \cite{Pincon2016}, we have to consider in the present case that the lifetime of the mode oscillation is of the order of $T_{\rm damp}$. 

The reasoning is as follows. Owing to the linearity of the wave equations, the displacement field at any time can be represented as the superposition of all the waves induced by each individual plume. Each of these waves has a finite lifetime, which is on the order of the damping timescale $T_{\rm damp}$.
Over the time interval $[-T/2,T/2]$, it thus appears obvious that the set of waves constituting $\vec{\xi}$ is generated by a finite number of plumes, denoted with $\mathcal{N}_T$. In the limit of $T\ggg  T_{\rm damp}$, as the plume emerging rate is constant and equal to $\mathcal{N}/\tau_{\rm p}$, with $\tau_{\rm p}$ the characteristic plume lifetime, the number of penetrating plumes that contributes to $\vec{\xi}$ in the time interval $[-T/2,T/2]$ is equal at leading order to $\mathcal{N}_T\sim \mathcal{N} \nu_{\rm p} T$, with $\nu_{\rm p}=1/\tau_{\rm p}$. Because of the incoherence of the convective plumes between each other, the wave velocity fields generated by different plumes negatively interfere. Assuming in addition that the excitation is a stationary process and the plumes are all identical and uniformly distributed over the sphere, the time and volume integrals in \eq{mean wave energy} can be written as the mode energy associated with the displacement field $\vec{\xi}_0(\vec{r},t;\theta_0,\varphi_0)$ that is generated by one single plume penetrating at $t_0=0$ in the solid angle $\dd \Omega_0=\sin \theta_0 \dd \theta_0 \dd \varphi_0$ and with the velocity $\vec{\mathcal{V}}_{{\rm p},0}$, multiplied by the number of penetrating plumes over $[-T/2,T/2]$, that is $\mathcal{N}_T\sim \mathcal{N} \nu_{\rm p} T$, and finally averaged over the plume angular position $(\theta_0,\varphi_0)$. Using the Parseval-Plancherel theorem in the time Fourier space, the mean mode energy can thus be expressed as
\algn{
\left\langle E \right\rangle=\mathcal{N}\frac{\nu_{\rm p}}{ 8 \pi^2 } \int_{V}\int_{\Omega_0}\left( \int_{-\infty}^{+\infty} 
\rho(r)\left| \widehat{\partial_t \vec{\xi}_0}(\vec{r},\omega;\theta_0,\varphi_0)\right|^2 \dd \omega\right) \dd\Omega_0 ~ \dd V \; ,
\label{mean energy ensemble}
}
where  the symbol ( $\hat{.}$ ) and $\omega$ denotes the time Fourier transform and the temporal angular frequency\footnote{The time Fourier transform of a function $X(\vec{r},t)$ is defined as\\$\mathcal{T}_{\rm F}[X]=\hat{X}(\vec{r},\omega)=\int_{-\infty}^{+\infty} X(\vec{r},t) ~ e^{-i\omega t} \dd t \; .$}. Equation~(\ref{mean energy ensemble}) can actually be retrieved through Eqs.~(5) and (12) of \cite{Pincon2016}, but integrated over the volume of the star in the present case.

To go further, the mode displacement field $\vec{\xi}_0$ is expanded onto the eigenfunctions of the \smash{$\vec{\mathcal{L}}^{\rm ad}$} operator, similarly to \eq{field expansion}, except that the instantaneous amplitude is induced only by the plume of velocity $\mathcal{V}_{{\rm p},0}$ and thus depends on $(\theta_0,\varphi_0)$; it is denoted with $a_{n\ell m}(t;\theta_0,\varphi_0)$.
Injecting such an expansion in \eq{mean energy ensemble} and using the orthogonality property of the adiabatic eigenfunctions, the mean total mode energy can be decomposed as
\algn{
\left \langle E \right  \rangle= \sum_{n=-\infty}^{+\infty} \sum_{\ell=0}^{+\infty} \sum_{m=-\ell}^{+\ell} \left\langle E_{n\ell m}\right\rangle \; ,
\label{E_nlm}
}
where $\left \langle E_{n\ell m} \right  \rangle$ is the mean oscillation energy of the harmonic $(n,\ell,m)$, that is,
\algn{
\langle E_{n\ell m}\rangle&= \mathcal{M}_{n\ell m}~ \mathcal{V}_{n\ell m}^2\; ,
\label{mean energy nlm}
}
with $\mathcal{V}_{n\ell m}$ the root mean square mode amplitude that reads
\begin{equation}
\mathcal{V}_{n\ell m}^2= \mathcal{N}\frac{\nu_{\rm p}}{ 8 \pi^2 } \int_{\Omega_0} \int_{-\infty}^{+\infty} 
\left| \widehat{ \partial_t a_{n\ell m}}(\omega;\theta_0,\varphi_0)\right|^2 \dd \omega \dd\Omega_0   \; .
\label{mean square velocity ensemble}
\end{equation}
To analytically express \eq{mean square velocity ensemble}, it is thus sufficient to compute the amplitude from \eqs{a_n solution}{a_n forced}~considering the model of plume velocity in \sectionname{}~\ref{plume model}.

\subsubsection{Asymptotic form with our plume model}
\label{asymptotic form}

According to \sectionname{}~\ref{plume model}, we assume that the mode driving takes place in the nearly adiabatically stratified penetration region of length $L_{\rm p}$ where the plumes are braked by buoyancy. 
Based on the asymptotic analysis of gravity modes by \cite{Shibahashi1979}, the WKB expressions of the radial eigenfunctions of frequency $\omega_{n\ell m}$, angular degree $\ell$ and azimuthal number $m$ in the penetration region are provided by Eq.~(B.24) of \cite{Pincon2016}, which corresponds to an evanescent wave up to a constant $A_{\rm c}$. In contrast, in the radiative zone, it is provided by the sum of Eqs.~(B.25)~and~(B.26), which corresponds to the sum of a progressive and a regressive wave up to a complex constant $A_{\rm r}$. The constants $A_{\rm r}$ and $A_{\rm c}$ have to be chosen in such a way to ensure the continuity of the radial and horizontal displacements at $r_{\rm b}$. This is met if $ |A_{\rm r}/A_{\rm c}| \approx (N_{\rm t}/\omega_{n\ell m})^{1/2}$ and $|\arg(A_{\rm r})|\approx \omega_{n\ell }/N_{\rm t}\ll 1$, in agreement with the more general computations of \cite{Lecoanet2013} or \cite{Pincon2016}.

First, using the WKB expressions of the eigenfunctions as described just before, \eqs{a_n solution}{a_n forced}{mode driving}{profil_Vp}, the mean square mode velocity in \eq{mean square velocity ensemble} can be rewritten, after some algebraic manipulations and the computation of the Fourier transform of the mode amplitude, as
\algn{
\mathcal{V}_{n\ell m}^2 \approx \frac{\mathcal{N}}{4 \eta_{n\ell m} \mathcal{M}_{n\ell m}^2}~ \frac{A_{\rm c}^2}{\omega_{n\ell m}} ~ \sqrt{\ell(\ell+1)}\mathcal{H}_\ell^2~\mathcal{B}_\ell~ \mathcal{C}_{n\ell m} \; ,
\label{V 1}
}
where $\mathcal{H}_\ell$ and $\mathcal{B}_\ell$ represent the radial and horizontal correlations between the plumes and the modes, whose expressions are provided by \smash{Eqs.~(32)-(34)} of \cite{Pincon2016}. The term $\mathcal{C}_{n\ell m}$ represents the temporal correlation that reads
\algn{
\mathcal{C}_{n\ell m}=\frac{\eta_{n \ell m} \nu_{\rm p} }{\pi} \int_{-\infty}^{+\infty} \frac{\omega \left | \widehat{f^2}(\omega)\right|^2}{(\omega-\omega_{n\ell m})^2+\eta_{n\ell m}^2} \dd\omega \; ,
\label{time correlation}
}
where we remind that $f(t)$ is the plume time evolution profile. We note that the constant $A_{\rm c}$ at the numerator of \eq{V 1} results from the inner product between the plume ram pressure and the eigenmode in the penetration region where the driving is assumed to be maximum (i.e., the volume integral in \eq{mode driving} is reduced to this region).
Second, using again the asymptotic expression of the adiabatic eigenfunctions, and considering only the contribution from the radiative core where most of the mode energy is contained, the mode mass is equal at leading order to \cite[e.g., see][for a similar computation]{Godart2009}
\algn{
\mathcal{M}_{n\ell m} \approx |A_{\rm r}|^2 \int_0^{r_{\rm b}} k_r \cos^2\left( \int_r^{r_{\rm b}} k_r \dd r\right) \dd r \approx |A_{\rm r}|^2\frac{n \pi}{2} \; ,
\label{mode mass}
}
where we have used the quantization condition $\int_0^{r_{\rm b}} k_r \dd r \approx n\pi $.
Finally, using \eqs{V 1}{mode mass}, the mean mode energy in \eq{mean energy nlm} becomes
\algn{
\left\langle E_{n\ell m}\right\rangle \approx \frac{\mathcal{N}}{2\eta_{n\ell m}}~\frac{\sqrt{\ell(\ell+1)}}{n\pi N_{\rm t}}~  \mathcal{H}_\ell^2\mathcal{B}_\ell \mathcal{C}_{n\ell m} \; ,
\label{mean E 1} 
}
where we have used the continuity condition $|A_{\rm c}/A_{\rm r}|^2=\omega_{n\ell m}/N_{\rm t}$. 
Before going further, it is also instructive to express in a simple way the global damping timescale in \eq{T_damp} using the asymptotic form of the eigenfunctions. As explained in \sectionname{}~\ref{globally quasi-adiabatic}, we assume that the contribution from the upper layers to the integral in \eq{T_damp} is negligible compared to the contribution from the radiative cavity, as suggested by the numerical computations of \cite{Belkacem2009}. Similarly to \eq{mode mass}, in the asymptotic regime, \smash{$\rho \left| \vec{\xi}_{n\ell m} \right|^2$ locally scales as $|A_{\rm r}|^2/\lambda_{n\ell m}$}, so that the global damping timescale can be reduced to the expression 
\algn{
T_{\rm damp}^{-1} \approx \dfrac{\int_0^{r_{\rm b}} t_{\rm damp}^{-1} \dd r/\lambda_{n\ell m}}{\int_0^{r_{\rm b}} \dd r/\lambda_{n\ell m}} \; .
\label{T_damp lambda 2}
}
%

\subsubsection{Temporal correlation $\mathcal{C}_{n\ell m}$}
\label{temporal correlation}

The temporal correlation between the oscillation modes and the plumes in \eq{time correlation} depends on the time evolution of the plume in the driving zone. Using both the convolution and Cauchy's residue theorems, we find in the exponential limiting case for $f=f_{\rm E}$
\algn{
\mathcal{C}_{n\ell m}^{\rm E}=\frac{8\nu_{\rm p}^2 ~\omega_{n\ell m}~(2\nu_{\rm p}+\eta_{n\ell m})}{\left[(2\nu_{\rm p}+\eta_{n\ell m})^2+\omega_{n\ell m}^2\right]^2} \; ,
\label{C_E}
}
and in the Gaussian limiting case for $f=f_{\rm G}$
\algn{
\mathcal{C}_{n\ell m}^{\rm G}=\pi \mathcal{I}_{\rm m}\left\{ ~\zeta e^{+\zeta^2}\left[ {\rm erf}(\zeta)-1\right]~\right\} \; ,
\label{C_G}
}
where $\zeta=(\eta_{n\ell m}-{\rm i} \omega_{n\ell m})/2\nu_{\rm p}$, $\mathcal{I}_{\rm m}()$ denotes the imaginary part, and ${\rm erf}()$ is the error function.
When $\eta_{n\ell m} \ll \nu_{\rm p}\ll \omega_{n\ell m}$, we have in a good approximation
\algn{
\mathcal{C}_{n\ell m}^{\rm E}\approx 16 \frac{\nu_{\rm p}^3}{\omega_{n\ell m}^3}  \; ,
\label{C_E approx}
}
which results from the contribution from the resonant frequencies such as $\omega\sim \omega_{n\ell m}$ to the integral in \eq{time correlation}, and

\algn{
\mathcal{C}_{n\ell m}^{\rm G}\approx \pi  \frac{\omega_{n\ell m}}{2 \nu_{\rm p}}~e^{-\omega_{n\ell m}^2/4\nu_{\rm p}^2}+\sqrt{\pi}\frac{8 \nu_{\rm p}^3}{\omega_{n\ell m}^3}\frac{ \eta_{n\ell m}}{2\nu_{\rm p}}\; ,
\label{C_G approx}
}
where the first term results from the contribution from the high frequencies such as $\omega\sim \omega_{n\ell m}$ to the integral in \eq{time correlation}, while the second term results from the contribution from the low frequencies such as $\omega \lesssim \nu_{\rm p}$.

\subsubsection{Damping rate}
\label{damping rate}

The expression of the damping rate $\eta_{n\ell m}$ is provided by \eq{eta_n}.
Actually, this expression is equivalent to the expression of the damping rate found by \cite{Godart2009} within the quasi-adiabatic and asymptotic limits. Using our notation, Eqs.~(14)~and~(15) of \cite{Godart2009} can be rewritten in the form \cite[see also Sect. 5.4.2 of][]{mathese}
\algn{
\eta_{n\ell m}\approx  \frac{[\ell(\ell+1)]^{3/2}}{2\omega_{n\ell m}^3} \frac{1}{n\pi } \int_0^{r_{\rm b}} \frac{H^2}{t_{\rm R}} N_T^2N \frac{\dd r}{r^3} \; ,
\label{Godart 1}
}
where $H$ is the temperature scale height and $N_T$ is the part of the Brunt-Väisälä frequency related to the temperature gradient only. According to \eqs{quantization}{k_r}, \eq{Godart 1} can be formulated as
\algn{
\eta_{n\ell m}\approx \frac{1}{2}\frac{\displaystyle \int_0^{r_{\rm b}} \dfrac{H^2}{t_{\rm R}} \frac{\ell(\ell+1) N_T^2}{\omega_{n\ell m}^2r^2} N \frac{\dd r}{r}}{\displaystyle\int_0^{r_{\rm b}} N \frac{\dd r}{r}}\lesssim  \frac{1}{2}\frac{\displaystyle \int_0^{r_{\rm b}} \dfrac{H^2}{t_{\rm R}} k_r^2 N \frac{\dd r}{r}}{\displaystyle\int_0^{r_{\rm b}} N \frac{\dd r}{r}}\; ,
\label{Godart 2}
}
where we used $N_T^2 \le N^2$ to write the inequality. Equation~(\ref{Godart 2}) shows that \smash{$\eta_{n\ell m} \propto \ell(\ell+1)/\omega_{n\ell m}^2$}. Moreover, the local damping timescale was defined as \smash{$t_{\rm damp} \sim (\lambda/H)^2~ t_{\rm R} \sim (k_rH)^{-2} t_{\rm R}$.} Therefore, according to \eq{T_damp lambda 2}, $\eta_{n\ell m}$ turns out to be $\mathcal{O}(1/T_{\rm damp})$. In addition, the comparison with the expression given by \eqs{T_damp}{mode coupling}{eta_n}~confirms a posteriori the scaling of the norm of the \smash{$\vec{\mathcal{L}}^{\rm nad}$} operator as $\omega_{n\ell m}^2$ and that of the \smash{$\mathcal{L}^{\rm nad1}$} operators as $H^2/\lambda^2$, such as assumed in \sectionname{}~\ref{scaling}.

\subsubsection{Simplified analytical expression}
\label{simplified}

For small angular degrees, that is $\ell \lesssim r_{\rm b} /b\sim 50$ for the Sun, which is sufficient for the present study, a very good analytical expression of $\mathcal{B}_\ell$ is provided by Eq.~(37) of \cite{Pincon2016}.
Moreover, within the large Péclet number regime, the length of the penetration zone is expected to be much smaller than the characteristic decay length of the mode toward the surface. We remind that the mode is evanescent in this region and that its decay length is equal to $r_{\rm b}/\sqrt{\ell(\ell+1)}$ \citep[e.g.,][]{Shibahashi1979}. As a consequence, the eigenfunctions slowly vary in the penetration region and the radial gradient of the plume ram pressure contained in $\mathcal{H}_\ell$ can be seen as a Dirac function, so that we have in a good approximation $\mathcal{H}_l^2\approx  r_{\rm b} \rho_{\rm b}  V_{\rm b}^4 $. \cite{Pincon2016} numerically demonstrated the validity of this simplification. Using both simplifications for $\mathcal{B}_\ell$ and $\mathcal{H}_\ell$, \eq{mean E 1} can be analytically expressed by \eq{luminosity approximate}.

\section{Non-adiabatic mode displacement basis}
\label{basis}

According to \cite{Chandra1964} or \cite{Unno1989}, the eigenfunctions of the $\vec{\mathcal{L}}^{\rm ad}$ operator in \eq{complete momentum eq} form, at each time, a complete basis of the displacement field over the stellar volume $V$ (i.e., beyond which the stellar density vanishes, the so-called zero-boundary conditions) when the oscillations are adiabatic. In this section, we show that this holds true in the non-adiabatic case.

To show this, limiting ourself to non-rotating stars, it is actually sufficient to demonstrate the hermicity of the \smash{$\vec{\mathcal{L}}^{\rm ad}$} operator. To do so, we adapt the demonstration done in the adiabatic case by \citet[][see Chaps. 14.2 and 14.3]{Unno1989} to the non-adiabatic case.
We consider two trial displacement fields, $\vec{\xi}$ and $\tilde{\vec{\xi}}$, that are solutions of the non-adiabatic oscillation equations and that are expanded onto the spherical harmonics basis. First, computing the dot product of \eq{complete momentum eq} with \smash{$\rho \tilde{\vec{\xi}}^\star$}, where the $(^\star)$ symbol denotes the complex conjugate, using the continuity equation in \eq{continuity eq}, and integrating over the stellar volume, we obtain
\algn{
\mathcal{I}&=\int_{V} \rho \tilde{\vec{\xi}}^\star \cdot \partial_t^2 \vec{\xi}\dd V = \int_{V} \left( -\rho \tilde{\vec{\xi}}^\star \cdot \vec{\mathcal{L}}^{\rm ad}\left( \vec{\xi}\right)+ \rho \tilde{\vec{\xi}}^\star \cdot \vec{F} \right) \dd V \nonumber \\
&-\int_{V} \vec{\nabla}\cdot \left( \Gamma_1 \varv_T p \frac{\delta S}{c_p} \tilde{\vec{\xi}}^\star\right) \dd V 
-\int_{V} \Gamma_1 \varv_T p \frac{\delta S}{c_p} \frac{\widetilde{\delta\rho}^{\star}}{\rho} \dd V \; ,
\label{I1}
}
where $\vec{F}$ represents a given forcing term. Second, computing the dot product of \eq{momentum eq} with \smash{$\rho\tilde{\vec{\xi}}^\star$}, and then proceeding as in Eq.~(14.18) of \cite{Unno1989}, i.e., expressing the left-hand side of the obtained expression in a flux-conservative form, using \eqs{continuity eq}{Poisson eq}, the hydrostatic equilibrium equation, as well as the equation of state in \eq{EOS eq} rewritten as
\algn{
\frac{\rho^\prime}{\rho}=\frac{p^\prime}{\rho c^2} + \xi_r \frac{N^2}{g} -\varv_T \frac{\delta S}{c_p} \; ,
}
with $N^2$ the square Brunt-Väisälä frequency provided in \eq{Brunt} and $\xi_r$ the radial displacement, we obtain after some algebraic manipulations
\algn{
\mathcal{I} &=-\int_{V} \left( \frac{\tilde{p}^{\prime\star} p^\prime}{\rho c^2}+N^2\rho \tilde{\xi}_r^\star \xi_r-\frac{\vec{\nabla}\tilde{\psi}^{\prime\star}\cdot\vec{\nabla}\psi^\prime}{4\pi G}- \rho \tilde{\vec{\xi}}^\star \cdot \vec{F} \right)\dd V\nonumber\\
&-\int_{V} \vec{\nabla}\cdot \left( p^\prime \tilde{\vec{\xi}}^\star +\rho \psi^\prime \tilde{\vec{\xi}}^\star\right) \dd V-\int_{V} \vec{\nabla}\cdot \left( \frac{\psi^\prime\vec{\nabla}\tilde{\psi}^{\prime\star}}{4\pi G}\right) \dd V \nonumber\\
&+\int_{V} \left( p^\prime \varv_T \frac{\widetilde{\delta S}^\star}{c_p}-\tilde{\xi}_r^\star \deriv{p}{r}\varv_T \frac{\delta S}{c_p} \right) \dd V \; .
\label{I2}
}
Owing to the zero-boundary conditions, the second integrals in \eqs{I1}{I2}~vanish in virtue of the Green-Ostrogradsky's theorem.
Therefore, equating \eq{I1} with \eq{I2} and replacing $\widetilde{\delta \rho}$ by \eq{EOS eq}, we find
\algn{
&\int_{V} \rho \tilde{\vec{\xi}}^\star \cdot \vec{\mathcal{L}}^{\rm ad}\left(\vec{\xi} \right) \dd V=\int_{V} \left( \frac{\tilde{p}^{\prime\star} p^\prime}{\rho c^2}+N^2\rho \tilde{\xi}_r^\star \xi_r-\frac{\vec{\nabla}\tilde{\psi}^{\prime\star}\cdot\vec{\nabla}\psi^\prime}{4\pi G}\right)\dd V\nonumber\\
 &-\sum_{\ell,m}\frac{(\ell+1)}{4\pi G} \left[r\psi^\prime_{\ell m}(r)\tilde{\psi}^{\prime\star}_{\ell m}(r)\right]_{r=R_V} \nonumber\\
& +\int_{V}\left(\Gamma_1\varv_T^2 p \frac{\widetilde{\delta S}^\star\delta S}{c_p^2}-\varv_T\left[p^\prime\frac{\widetilde{\delta S}^\star}{c_p}+\frac{\delta S}{c_p} \tilde{p}^{\prime\star}\right]\right)\dd V\; ,
\label{xi dot Lxi}
}
where $\psi_{\ell m}$ are the radial functions of the perturbation of the gravitational potential according to the expansion on the orthonormal spherical harmonics, and $R_{V}$ is the radius of the sphere of volume $V$. The first integral and the sum in the right-hand side of \eq{xi dot Lxi} correspond to the expression found by \cite{Unno1989} within the adiabatic hypothesis, the last term of which results from the zero-boundary conditions at $r=R_{V}$, i.e., $(\dd \psi_{\ell m}/\dd r) = -(\ell +1) \psi_{\ell m} / r$ according to the Poisson's equation with a null density. In contrast, the last integral in the right-hand side of \eq{xi dot Lxi} results from the non-adiabatic effects. Despite of this difference, \eq{xi dot Lxi} remains symmetric with respect to \smash{$\tilde{\vec{\xi}}^\star$} and $\vec{\xi}$, i.e.,
\algn{
\int_{V} \rho \tilde{\vec{\xi}}^\star \cdot \vec{\mathcal{L}}^{\rm ad}\left(\vec{\xi} \right) \dd V=
\int_{V} \rho  \vec{\mathcal{L}}^{\rm ad}\left(\tilde{\vec{\xi}}^\star\right)\cdot \vec{\xi}  \dd V \; .
\label{hermicity}
}
Therefore, $\vec{\mathcal{L}}^{\rm ad}$ is Hermitian in the non-adiabatic case too. According to the spectral theorem, the set of the eigenfunctions of \smash{$\vec{\mathcal{L}}^{\rm ad}$} also forms a basis of the displacement field over $V$ in the non-adiabatic case. From \eq{hermicity} and the eigenvalue relation in \eq{eigenvalue}, it is straightforward to demonstrate that these eigenfunctions are orthogonal.

\section{Mean mode radial velocity}
\label{mean velocity}

In this section, we briefly summarize the computation of the mean apparent surface velocity following the Appendix C of \cite{Belkacem2009}.

First, we remind that in the slow rotation limit, which is valid for the considered frequency range in the Sun, the eigenfrequencies are slightly shifted as a function of $\ell$ and $m$ around the value predicted in the non-rotating case \citep[e.g.,][]{Ledoux1951}. The oscillation power spectrum of the solar gravity modes is thus composed of a forest of peaks, each of them associated with a tuple $(n,\ell,m)$. It is thus necessary to compute the mean apparent radial velocity for all these components. However, for the sake of simplicity, the computation is made at first approximation by neglecting the effect of rotation on the spatial shape of the mode. Therefore, we assume the total mode displacement field $\vec{\xi}$ can still be expanded onto the eigenfunctions computed in the non-rotating case, that is,
\algn{
\vec{\xi}(\vec{r},t)=\sum_{n=-\infty}^{+\infty}\sum_{\ell=0}^{+\infty} \sum_{m=-\ell}^{+\ell} \tilde{a}_{n\ell m}(t)~ \vec{\xi}_{n\ell m} (r,\theta,\varphi) \;,
\label{total field expansion}
}
where $\tilde{a}_{n\ell m}(t)$ is an instantaneous amplitude, $\vec{\xi}_{n\ell m}$ is normalized by the value of the radial displacement at the photosphere, and $(r,\theta ,\varphi)$ are the spherical coordinate system in the observer's frame such as the polar axis (i.e., $\theta=0$) corresponds to the direction of the stellar rotation axis and the origin (i.e, $r=0$) corresponds to the stellar center. Under this approximation, the velocity component associated with the tuple $(n,\ell, m)$ is thus merely equal to 
\algn{
\vec{\varv}_{n\ell m}(\vec{r},t)=\partial_t \tilde{a}_{n\ell m}(t)~\vec{\xi}_{n\ell m} (\vec{r})\; .
\label{velocity nlm}
}

Second, we define a spherical coordinate system $(r,\Theta,\Phi)$ in the observer's frame whose origin (i.e., $r=0$) corresponds to the star center and the polar axis (i.e., $\Theta=0$) is directed toward the observer. Given the large distance from the observer, the direction of the line-of-sight can be considered as equal to the unit vector $\vec{n}$ parallel to the polar axis. Moreover, owing to the small amplitudes of the oscillations near the stellar surface, the shell in which the considered absorption line forms remains undeformed at first order. Within this framework, each infinitesimal surface element $\dd^2 S_{\rm abs}$ of the absorption shell emits a number proportional to $\dd^2 N_{\lambda}=h(\mu)~\mu ~\dd^2 S_{\rm abs}$ of photons at wavelength $\lambda$ and per unit of time toward the observer, where $\mu=\cos \Theta$, $h(\mu)$ is the limb-darkening function, $\dd^2 S_{\rm abs}=r_{\rm abs}^2~\dd \mu ~\dd \Phi$, and $r_{\rm abs}$ is the radius of the considered shell.
The mean radial velocity is therefore defined as the average of the radial velocity over the disk weighted by the number of photons received by the observer from each point of its surface, that is \citep[e.g.][]{Dziembowski1977},
\algn{
\varv_{n\ell m}^{\rm rad}=\dfrac{\int_0^1\int_0^{2\pi} \left[\vec{\varv}_{n\ell m}(r_{\rm abs},\Theta,\Phi,t)\cdot\vec{n}\right]h(\mu)\mu r_{\rm abs}^2\dd \mu \dd \Phi}{\int_0^1 \int_0^{2\pi}h(\mu)\mu r_{\rm abs}^2\dd \mu \dd \Phi} \; .
\label{radial velocity}
}

To go a step further, the mean apparent velocity is defined as the statistical or, equivalently for an ergodic process, time average quantity
\algn{
\varv_{n\ell m}^{\rm app}= \sqrt{\langle \varv_{n\ell m}^{\rm rad}{}^2 \rangle} \; .
\label{apparent velocity}
}
Using \eqs{velocity nlm}{radial velocity}~and following all the derivation steps in Appendix C of \cite{Belkacem2009},
it is straightforward to show that \eq{apparent velocity} is equal to
\algn{
\varv_{n\ell m}^{\rm app}=\sqrt{\left\langle \left(\partial_t \tilde{a}_{n\ell m}\right)^2 \right\rangle} \left|\alpha_\ell^m\xi_{n\ell m}^r(r_{\rm abs})+\beta_\ell^m \xi_{n\ell m}^h(r_{\rm abs}) \right| \; ,
\label{v_app}
}
with
\algn{
\alpha_{\ell }^m&=N_\ell^m~ \left|P_\ell^m(\cos \Theta_0) \right|~u_\ell \label{alpha_lm}\\
\beta_{\ell }^m&=N_\ell^m ~\left|P_\ell^m(\cos \Theta_0)) \right|~ \varv_\ell \label{beta_lm}\; ,
}
where $\Theta_0$ is the angle between the rotation axis and the line-of-sight, $P_\ell^m(\mu)$ is the associated Legendre polynomial, and $N_\ell^m$, $u_\ell$ and $\varv_\ell$ are defined by 
\algn{
N_{\ell}^m&=\sqrt{\frac{2\ell+1}{4\pi}} \sqrt{\frac{(\ell -m)!}{(\ell+m)!}}\\
u_\ell&=\int_{0}^1 \mu^2 \tilde{h}(\mu)P_\ell(\mu)\dd \mu\\
\varv_\ell&= \ell\int_{0}^1 \mu \tilde{h}(\mu)\left[P_{\ell-1}(\mu) -\mu P_\ell(\mu)\right]\dd \mu\; ,
}
in which the $\tilde{h}(\mu)$ function must be replaced by the properly normalized limb darkening function, that is,
\algn{
\tilde{h}(\mu)=\dfrac{h(\mu)}{\int_0^1 h(\mu) \mu \dd \mu} \; .
}
Finally, according to \eq{total field expansion} and the orthogonality of the eigenfunctions, it is straightforward to show that the mean square amplitude is provided by
\algn{
\left\langle \left(\partial_t \tilde{a}_{n\ell m}\right)^2 \right\rangle=\frac{\left\langle E_{n\ell m}\right\rangle}{\mathcal{M}_{n\ell m}} \; ,
}
where $\left\langle E_{n\ell m}\right\rangle$ is the mean mode energy for the tuple $(n,\ell,m)$ given in \eq{mean energy nlm} and $\mathcal{M}_{n\ell m}$ is the mode mass in \eq{orthogonality}. Therefore, the apparent velocity in \eq{v_app} can be rewritten as in \eqs{v_app_2}{visibility}.

\section{GOLF detection threshold}
\label{threshold_app}

According to \cite{Appourchaux2000}, the threshold signal-to-noise ratio above which a peak over a frequency interval $\Delta f$ in the power spectral density (PSD) can be considered as a statistically-relevant signal is equal to
\algn{
\frac{s_{\rm th}}{\tilde{s}} \approx \ln\left(\frac{T  \Delta f}{ p_{\rm th} }\right)\; ,
\label{s_th}
}
where $\tilde{s}$ is the mean noise level over $\Delta f$ and $p_{\rm th}$ is the false alarm probability that the measurement is due to pure noise. We note that the frequency interval $\Delta f$ in \eq{s_th} must be chosen small enough for the PSD to remain about constant over this range but large enough for the number of frequency bins inside to be high enough.
The mean noise level between $10~\mu$Hz and $100~\mu$Hz can be estimated in a simple way, for example, through the analysis of the 10-years GOLF data of \citet[][see Fig. S1.]{Garcia2007}, i.e.,
\algn{
\tilde{s} \sim 3~10^{3} ~\left( \frac{\nu }{25~\mu{\rm Hz}}\right)^{-3/4}~\mbox{m}^2~\mbox{s}^{-2} ~\mbox{Hz}^{-1}\; .
\label{mean noise}
}
The value of the PSD at the level of detection, i.e. $s_{\rm th}$, can then be converted into a threshold value $\varv_{\rm th}$ for the root mean square velocity by integrating the PSD over one frequency bin of width $1/T$ and computing the square root of the result, leading to \eq{v_th}. We emphasize that the expression of the mean noise level provided in \eq{mean noise} is sufficient for our purpose since the threshold velocity depends on $\tilde{s}$ to the power $1/2$ only, and thus is little impacted by the error made with our simple estimate of the magnitude and the frequency exponent in this expression.

\end{document}